\documentclass[ALICE,manyauthors]{cernphprep}
\usepackage[comma,square,numbers,sort&compress]{natbib}
\usepackage{hyperref}
\usepackage{lineno}
\usepackage{xspace}
\usepackage{bm}
\usepackage{xcolor}
\usepackage{graphicx,subfigure}
\usepackage[inline]{enumitem}
\usepackage{nicematrix}\usepackage{multirow}

\usepackage{upgreek}
\usepackage[T1]{fontenc}
\usepackage{orcidlink}

\begin{document}
%

\newcommand{\pp}           {pp\xspace}
\newcommand{\ppbar}        {\mbox{$\mathrm {p\overline{p}}$}\xspace}
\newcommand{\XeXe}         {\mbox{Xe--Xe}\xspace}
\newcommand{\PbPb}         {\mbox{Pb--Pb}\xspace}
\newcommand{\pA}           {\mbox{pA}\xspace}
\newcommand{\pPb}          {\mbox{p--Pb}\xspace}
\newcommand{\AuAu}         {\mbox{Au--Au}\xspace}
\newcommand{\dAu}          {\mbox{d--Au}\xspace}

\newcommand{\s}            {\ensuremath{\sqrt{s}}\xspace}
\newcommand{\snn}          {\ensuremath{\sqrt{s_{\mathrm{NN}}}}\xspace}
\newcommand{\pt}           {\ensuremath{p_{\textrm T}}\xspace}
\newcommand{\meanpt}       {$\langle p_{\mathrm{T}}\rangle$\xspace}
\newcommand{\ycms}         {\ensuremath{y_{\textrm CMS}}\xspace}
\newcommand{\ylab}         {\ensuremath{y_{\textrm lab}}\xspace}
\newcommand{\etarange}[1]  {\mbox{$\left | \eta \right |~<~#1$}}
\newcommand{\yrange}[1]    {\mbox{$\left | y \right |~<~#1$}}
\newcommand{\dndy}         {\ensuremath{\mathrm{d}N_\mathrm{ch}/\mathrm{d}y}\xspace}
\newcommand{\dndeta}       {\ensuremath{\mathrm{d}N_\mathrm{ch}/\mathrm{d}\eta}\xspace}
\newcommand{\avdndeta}     {\ensuremath{\langle\dndeta\rangle}\xspace}
\newcommand{\dNdy}         {\ensuremath{\mathrm{d}N_\mathrm{ch}/\mathrm{d}y}\xspace}
\newcommand{\Npart}        {\ensuremath{N_\mathrm{part}}\xspace}
\newcommand{\Ncoll}        {\ensuremath{N_\mathrm{coll}}\xspace}
\newcommand{\dEdx}         {\ensuremath{\textrm{d}E/\textrm{d}x}\xspace}
\newcommand{\RpPb}         {\ensuremath{R_{\textrm pPb}}\xspace}

\newcommand{\onepointeight}{$\sqrt{s}~=~1.8$~Te\kern-.1emV\xspace}
\newcommand{\sixthreezero} {$\sqrt{s}~=~630$~Ge\kern-.1emV\xspace}
\newcommand{\twoH}         {$\sqrt{s}~=~200$~Ge\kern-.1emV\xspace}
\newcommand{\nineH}        {$\sqrt{s}~=~0.9$~Te\kern-.1emV\xspace}
\newcommand{\seven}        {$\sqrt{s}~=~7$~Te\kern-.1emV\xspace}
\newcommand{\twosevensix}  {$\sqrt{s}~=~2.76$~Te\kern-.1emV\xspace}
\newcommand{\five}         {$\sqrt{s}~=~5.02$~Te\kern-.1emV\xspace}
\newcommand{\thirteen}      {$\sqrt{s}~=~13$~Te\kern-.1emV\xspace}
\newcommand{\twosevensixnn}{$\sqrt{s_{\mathrm{NN}}}~=~2.76$~Te\kern-.1emV\xspace}
\newcommand{\fivenn}       {$\sqrt{s_{\mathrm{NN}}}~=~5.02$~Te\kern-.1emV\xspace}
\newcommand{\LT}           {L{\'e}vy-Tsallis\xspace}
\newcommand{\GeVc}         {\ensuremath{\mathrm{Ge\kern-.1emV}/c}\xspace}
\newcommand{\MeVc}         {\ensuremath{\mathrm{Me\kern-.1emV}/c}\xspace}
\newcommand{\TeV}          {Te\kern-.1emV\xspace}
\newcommand{\GeV}          {Ge\kern-.1emV\xspace}
\newcommand{\MeV}          {Me\kern-.1emV\xspace}
\newcommand{\GeVmass}      {Ge\kern-.2emV/$c^2$\xspace}
\newcommand{\MeVmass}      {Me\kern-.2emV/$c^2$\xspace}
\newcommand{\lumi}         {\ensuremath{\mathcal{L}}\xspace}

\newcommand{\ITS}          {\textrm{ITS}\xspace}
\newcommand{\TOF}          {\textrm{TOF}\xspace}
\newcommand{\ZDC}          {\textrm{ZDC}\xspace}
\newcommand{\ZDCs}         {\textrm{ZDCs}\xspace}
\newcommand{\ZNA}          {\textrm{ZNA}\xspace}
\newcommand{\ZNC}          {\textrm{ZNC}\xspace}
\newcommand{\SPD}          {\textrm{SPD}\xspace}
\newcommand{\SDD}          {\textrm{SDD}\xspace}
\newcommand{\SSD}          {\textrm{SSD}\xspace}
\newcommand{\TPC}          {\textrm{TPC}\xspace}
\newcommand{\TRD}          {\textrm{TRD}\xspace}
\newcommand{\VZERO}        {\textrm{V0}\xspace}
\newcommand{\VZEROA}       {\textrm{V0A}\xspace}
\newcommand{\VZEROC}       {\textrm{V0C}\xspace}
\newcommand{\Vdecay} 	   {\ensuremath{V^{0}}\xspace}

\newcommand{\ee}           {\ensuremath{e^{+}e^{-}}} 
\newcommand{\pion}         {\ensuremath{\uppi}\xspace}
\newcommand{\kaon}         {\ensuremath{\textrm{K}}\xspace}
\newcommand{\pip}          {\ensuremath{\uppi^{+}}\xspace}
\newcommand{\pim}          {\ensuremath{\uppi^{-}}\xspace}
\newcommand{\kap}          {\ensuremath{\textrm{K}^{+}}\xspace}
\newcommand{\kam}          {\ensuremath{\textrm{K}^{-}}\xspace}
\newcommand{\pbar}         {\ensuremath{\textrm\overline{p}}\xspace}
\newcommand{\kzero}        {\ensuremath{{\textrm K}^{0}_{\textrm{S}}}\xspace}
\newcommand{\lmb}          {\ensuremath{\Lambda}\xspace}
\newcommand{\almb}         {\ensuremath{\overline{\Lambda}}\xspace}
\newcommand{\Om}           {\ensuremath{\Omega^-}\xspace}
\newcommand{\Mo}           {\ensuremath{\overline{\Omega}^+}\xspace}
\newcommand{\X}            {\ensuremath{\Xi^-}\xspace}
\newcommand{\Ix}           {\ensuremath{\overline{\Xi}^+}\xspace}
\newcommand{\Xis}          {\ensuremath{\Xi^{\pm}}\xspace}
\newcommand{\Oms}          {\ensuremath{\Omega^{\pm}}\xspace}
\newcommand{\degree}       {\ensuremath{^{\textrm o}}\xspace}

\newcommand{\kamp}          {\ensuremath{\textrm{K}^{\mp}}\xspace}
\newcommand{\pipm}          {\ensuremath{\uppi^{\pm}}\xspace}

\newcommand{\jpsi}{\textrm J/$\psi$}
\newcommand{\psip}{$\psi^\prime$}
\newcommand{\jpsiDY}{\textrm J/$\psi$\,/\,DY}
\newcommand{\chic}{$\chi_{\textrm c}$}
\newcommand{\pizero}{$\pi^{0}$}
\newcommand{\ccbar}{\ensuremath{\mathrm{c\overline{c}}}}
\newcommand{\bbbar}{\ensuremath{\mathrm{b\overline{b}}}}
\newcommand{\Dzero}{\ensuremath{\mathrm{D^{0}}}\xspace}
\newcommand{\Dzerobar}{$\overline{\Dzero}$\xspace}
\newcommand{\Dpm}{\ensuremath{\mathrm{D^{\pm}}}}
\newcommand{\Ds}{\ensuremath{\mathrm{D_{s}^{\pm}}}}
\newcommand{\Dstar}{\ensuremath{\mathrm{D^{*\pm}}}\xspace}


\newcommand{\ezdc}{$E_{\textrm ZDC}$}

\newcommand{\GeVcsq}{\ensuremath{\mathrm{Ge\kern-.1emV}/c^2}\xspace}
\newcommand{\zch}{\ensuremath{z_{||}^{\textrm{ch}}}\xspace}
\newcommand{\zchdet}{\ensuremath{z_{||}^{\textrm{ch,det}}}\xspace}
\newcommand{\zchgen}{\ensuremath{z_{||}^{\textrm{ch,part}}}\xspace}
\newcommand{\y}{\ensuremath{y}}
\newcommand{\dedx}{d$E$/d$x$}
\newcommand{\dndydpt}{${\textrm d}^2N/({\textrm d}y {\textrm d}\pt)$}
\newcommand{\dndetadpt}{\ensuremath{{\textrm d}^2N/({\textrm d}\eta {\textrm d}\pt)}}
\newcommand{\dsigmadetadpt}{\ensuremath{{\textrm d}^2\sigma/({\textrm d}\eta {\textrm d}\pt)}}
\newcommand{\dndpt}{d$N$/d\pt}
\newcommand{\zpar}{\ensuremath{z_{||}}\xspace}
\newcommand{\zpargen}{\ensuremath{z_{||}^{\mathrm{part}}}\xspace}
\newcommand{\zpardet}{\ensuremath{z_{||}^{\mathrm{det}}}\xspace}
\newcommand{\ptchjet}{\ensuremath{p_{\mathrm{T,ch\,jet}}}\xspace}
\newcommand{\pchjet}{\ensuremath{\vec{p}_{\mathrm{ch\, jet}}}\xspace}
\newcommand{\pjet}{\ensuremath{\vec{p}_{\mathrm{jet}}}\xspace}
\newcommand{\ptjet}{\ensuremath{p_{\mathrm{T,jet}}}\xspace}
\newcommand{\ptchjetgen}{\ensuremath{p_{\mathrm{T,ch\,jet}}^{\mathrm{part}}}\xspace}
\newcommand{\ptchjetcorr}{\ensuremath{p_{\mathrm{T,ch\,jet}}^{\mathrm{corr}}}\xspace}
\newcommand{\ptchjetraw}{\ensuremath{p_{\mathrm{T,ch\,jet}}^{\mathrm{raw}}}\xspace}
\newcommand{\ptchjetdet}{\ensuremath{p_{\mathrm{T,ch\,jet}}^{\mathrm{det}}}\xspace}
\newcommand{\ptd}{\ensuremath{p_{\mathrm{T,D^0}}}\xspace}
\newcommand{\pd}{\ensuremath{\vec{p}_{\mathrm{D}}}\xspace}
\newcommand{\pdzero}{\ensuremath{\vec{p}_{\mathrm{D^{0}}}}\xspace}
\newcommand{\ptdgen}{\ensuremath{p_{\mathrm{T,D}}^{\mathrm{part}}}\xspace}
\newcommand{\ptddet}{\ensuremath{p_{\mathrm{T,D}}^{\mathrm{det}}}\xspace}
\newcommand{\antikt}{anti-\ensuremath{k_{\mathrm{T}}}\xspace}
\newcommand{\kt}{\ensuremath{k_{\mathrm{T}}}}
\newcommand{\pthard}{\ensuremath{p_{\mathrm{T,hard}}}}
\newcommand{\etajet}{\ensuremath{\eta_{\mathrm{jet}}}}
\newcommand{\phijet}{\ensuremath{\phi_{\mathrm{jet}}}}
\newcommand{\deltapt}{\ensuremath{\delta p_{\textrm T}}}

\begin{titlepage}
\PHyear{2022}       
\PHnumber{070}      
\PHdate{29 March}  

    \title{Measurement of the production of charm jets tagged with \Dzero\ mesons in pp collisions at $\sqrt{s}$ = 5.02 and 13 TeV}
\ShortTitle{ \Dzero-jet production in pp collisions at $\sqrt{s}$ = 5.02 and 13 TeV}   

\Collaboration{ALICE Collaboration\thanks{See Appendix~\ref{app:collab} for the list of collaboration members}}
\ShortAuthor{ALICE Collaboration} 

\begin{abstract}
The measurement of the production of charm jets, identified by the presence of a \Dzero\ meson in the jet constituents, is presented in proton--proton collisions at centre-of-mass energies of $\sqrt{s}~=$~5.02 and 13~TeV with the ALICE detector at the CERN LHC.
The \Dzero\ mesons were reconstructed from their hadronic decay \Dzero $\rightarrow$\ \kam \pip\ and 
the respective charge conjugate.
Jets were reconstructed from \Dzero-meson candidates and charged particles using the anti-$k_\text{T}$ algorithm, in the jet transverse momentum range $5 < \ptchjet < 50$ \GeVc, pseudorapidity $\left | \etajet \right | < 0.9-R$, and with the jet resolution parameters $R = 0.2, 0.4, 0.6$. 
The distribution of the jet momentum fraction carried by a \Dzero\ meson along the jet axis (\zch) was measured in the range $0.4< \zch <1.0$ in four ranges of the jet transverse momentum.
Comparisons of results for different collision energies and jet resolution parameters are also presented. 
The measurements are compared to predictions 
from Monte Carlo event generators based on leading-order and next-to-leading-order perturbative quantum chromodynamics calculations. A generally good description of the main features of the data is obtained in spite of a few discrepancies at low \ptchjet.
Measurements were also done for $R=0.3$ at \five and are shown along with their comparisons to theoretical predictions in an appendix to this paper.

\end{abstract}
\end{titlepage}

\setcounter{page}{2} 

\section{Introduction}

In high-energy proton--proton (\pp) collisions, heavy quarks (charm and beauty) are produced in hard scatterings between the partons of the incoming protons. 
Since their masses are 
greater than the quantum chromodynamics (QCD) non-perturbative scale $\Lambda_{\textrm{QCD}}$, the production cross section of heavy quarks can be calculated using perturbative QCD (pQCD) methods~\cite{Cacciari:2005rk,Kniehl:2012ti,Maciula:2013wg,Cacciari:1998it,Benzke:2017yjn,Helenius:2018uul}. 
For example, the  Fixed-Order-Next-to-Leading-Logarithm (FONLL)~\cite{Cacciari:1998it} and General-Mass Variable-Flavor-Number Scheme (GM-VFNS)~\cite{Benzke:2017yjn,Helenius:2018uul} pQCD calculations can describe measurements of heavy-flavour meson production in \pp collisions at RHIC and LHC energies and the \ppbar collision data at the SPS and Tevatron~\cite{UA1:1986xzz, CDF:2009bqp, Adamczyk:2012af, PHENIX:2017ztp,ALICE:2021mgk, ATLAS:2015igt, CMS:2019uws, Aaij:2015bpa}.

Measurements of the production and substructure properties of heavy-flavour tagged jets provide additional information to that given by heavy-flavour hadron production. They offer a different sensitivity to study heavy-quark production processes and the contribution from higher-order processes, like gluon splitting and flavour excitation, which is useful to test pQCD calculations and tune Monte Carlo (MC) event generators~\cite{Anderle:2017cgl,Li:2021gjw}.
The transverse-momentum (\pt) differential production cross sections of charm and beauty jets were measured at the LHC in \pp collisions~\cite{Sirunyan:2016fcs,Acharya:2019zup,Chatrchyan:2012dk,Aad:2011td,ATLAS:2021agf,ALICE:2021wct} and were found to be consistent with next-to-leading order (NLO) pQCD calculations.
Further insight into heavy-quark production can be obtained through measurements of fully reconstructed heavy-flavour hadrons inside jets and studies of the jet momentum (or energy) fraction carried by the heavy-flavour hadron, \zpar, along the jet axis direction.
Studies of charm jets, containing \Dstar mesons, were performed in \pp collisions at RHIC at centre-of-mass energy \twoH by STAR~\cite{Abelev:2009aj}, CERN SPS at \sixthreezero by UA1~\cite{Albajar:1990wn}, LHC at \seven by ATLAS~\cite{Aad:2011td}, and in \ppbar collisions at Tevatron at \onepointeight by CDF~\cite{Abe:1989tc}. 
These measurements showed that the \zpar distribution is peaked at low \zpar values. The STAR low-\zpar enhancement cannot be described by event generators that include the leading-order charm-pair creation process ($\rm{gg/q\bar{q}} \rightarrow \rm{c\bar{c}}$) only. 
Also, the shape of the \zpar distributions measured by ATLAS is in disagreement with predictions from various Monte Carlo event generators at small values of \zpar. 
On the other hand, the \zpar distributions at \seven measured by the ALICE Collaboration are in good agreement with next-to-leading order pQCD calculations and different Monte Carlo event generator predictions~\cite{Acharya:2019zup}. 
These observations showed the need of further model refinements and suggest the importance of the contribution of higher-order processes to charm-quark production, e.g.~the ATLAS data can be described by enhancing the gluon-to-D meson fragmentation function~\cite{Chien:2015ctp}. 
It should be noted that ALICE and ATLAS measurements used different experimental methods and correction techniques as discussed in ~\cite{Acharya:2019zup}.

The heavy-flavour hadron in-jet fragmentation data can also help in constraining the gluon fragmentation functions (FFs). 
The FFs are usually assumed to be universal and are constrained from semi-inclusive electron--positron annihilation (SIA) data~\cite{ZEUS:2008axw,Belle:2005mtx}. 
The ATLAS measurement of the jet momentum fraction carried by \Dstar mesons~\cite{Aad:2011td} proved to be an important ingredient (together with SIA and the inclusive hadron production data) in the global fit analysis based on the Zero-Mass Variable-Flavor-Number Scheme (ZM-VFNS)~\cite{Anderle:2017cgl}. 
The new ALICE results presented in this paper, from two collision energies and for lower transverse momentum ranges, provide a valuable complementary input to this global fit analysis. 

Furthermore, understanding the heavy-flavour jet production in \pp collisions is crucial for the interpretation of results from collisions of heavy nuclei~\cite{Busza:2018rrf}. 
Lattice QCD calculations~\cite{HotQCD:2018pds, Borsanyi:2020fev,Satz_2006} predict that in ultra-relativistic heavy-ion collisions, a state of matter known as the quark--gluon plasma (QGP), where quarks and gluons are deconfined, can be produced~\cite{Heinz:2000bk, PHENIX:2004vcz}.
Heavy quarks are dominantly produced in hard scatterings at the initial stage of a collision, before the QGP formation, and their thermal production in the QGP is negligible. 
They traverse the medium and lose part of their energy via collisional and radiative processes~\cite{Aichelin:2012ww}.
Therefore, heavy quarks are ideal tomographic probes~\cite{Dong:2019byy} of the QGP~\cite{ALICE:2012ab,ALICE:2015ccw,ALICE:2018lyv, STAR:2018zdy, Sirunyan:2017xss}, allowing extraction of the medium transport properties~\cite{Rapp:2018qla, Cao:2018ews, STAR:2017kkh, ALICE:2020iug, ALICE:2021kfc}.
Studies of jets including heavy-flavour hadrons in such collisions can set additional constraints on the heavy-quark energy loss mechanism and the medium properties as they provide insight into how the lost energy is radiated and dissipated in the medium. 

In this paper, ALICE results on track-based jets (i.e. reconstructed using charged-particle constituents), tagged with the presence of a fully reconstructed \Dzero meson (\Dzero jet), in \pp collisions at \five and \thirteen at midrapidity are presented.
Thus, the \zpar variable associated with charged tracks is better denoted as \zch. 
These measurements extend the previous ALICE charm jet studies in pp collisions at \seven~\cite{Acharya:2019zup}. 
The better precision obtained with these new data samples allowed more differential \mbox{\Dzero-jet} studies to be conducted as a function of the jet resolution parameter ($R$) and to measure the \zch distributions in a larger number of charged-jet transverse momentum \ptchjet intervals.
The \Dzero-jet \ptchjet-differential cross sections and \zch distributions are reported in several \ptchjet ranges between 5 and 50~\GeVc and for $R =$~0.2, 0.4, 0.6. The \zch variable is defined in this article as
\begin{equation}
    \zch = \frac{\pchjet\cdot\pdzero}{\pchjet\cdot\pchjet},
    \label{Eq:Zpar}
\end{equation}
where $\pdzero$ is the total \Dzero-meson momentum and $\pchjet$ is the 
total track-based jet momentum.
Ratios of the \ptchjet cross sections obtained for the two energies and with different $R$ values are also presented. The $R$ dependence is sensitive to both perturbative and non-perturbative physics of the jet production and fragmentation, and provides information on the parton shower development~\cite{Dasgupta:2016bnd}.
The results are compared to predictions of the Monte Carlo PYTHIA~8.2~\cite{Pythia8} event generator and NLO pQCD POWHEG~\cite{Frixione:2007vw, Alioli:2010xd} calculations, matched to the PYTHIA~8 parton shower.

This paper is organised as follows: Section~\ref{Sec:detectoranddata} describes the ALICE detector and the utilised data samples, 
Sections~\ref{Sec:analysis} and~\ref{Sec:sys} provide details on the analysis procedure and the systematic uncertainties, respectively. 
Section~\ref{Sec:results} presents the final results compared to different model predictions. Finally, conclusions are given in Section~\ref{sec:summary}.

\section{Detector and data sample}\label{Sec:detectoranddata}

The reconstruction of heavy-flavour hadrons and charged jets in this analysis is done with 
three detectors located within a large solenoid in the central barrel of the \mbox{ALICE} experimental setup~\cite{alice-detector}: the Time Projection Chamber (\TPC), the Inner Tracking System (\ITS), and the Time-Of-Flight detector (\TOF). The TPC is a gaseous drift chamber detector used for track reconstruction and particle identification (PID), thanks to the measurement of the specific energy loss of particles in the detector gas due to ionisation. 
The ITS, a six-layer cylindrical silicon detector, complements the track reconstruction in the TPC, 
allowing for a precise determination of particle trajectories 
in the vicinity of the collision point and 
 the identification of charm-hadron decay vertices displaced by tens-to-hundreds of microns from the collision point. 
The resolution on the track impact parameter in the transverse plane to the primary vertex is better than 75 $\upmu$m for tracks with $\pt>1$~\GeVc~\cite{Abelev:2014ffa}.
Furthermore, the effectiveness of pion/kaon separation is enhanced by the TOF, a multi-gap resistive plate chamber detector, providing the time of flight of particles from the interaction point. 
Due to the low magnetic field (0.5 T), \mbox{ALICE} is capable of 
reconstructing low-momentum particles down to $\pt$ lower than 150~\MeVc.
The central barrel detectors cover the pseudorapidity range $|\eta|<0.9$ and full azimuthal angle of $\varphi \in [0, 2\pi]$. 
To provide uniform pseudorapidity acceptance, only those events having a primary vertex within $\pm10$ cm from the nominal beam collision position along the beam direction were analysed.

Events with the least possible bias were selected using a minimum bias trigger which helped in identifying beam--beam collisions by requiring an event to be accepted only if a signal was found in both scintillator arrays of the V0 detector covering the pseudorapidity intervals $-3.7<\eta<-1.7$ and $2.8<\eta<5.1$.
The V0 detector was used in combination with the Silicon Pixel Detector (SPD), which comprises the first two layers of the ITS, to reduce the background due to beam--gas interactions. 
In addition, a dedicated algorithm based on multiple-vertex searches 
in the SPD was used in order to reduce pile-up events containing two or more primary vertices.

The data samples analysed in this paper consist of $0.99 \times 10^9$ minimum bias events from pp collisions at \five recorded in 2017, corresponding to an integrated luminosity of $\mathcal{L}_{int}=(19.3\pm0.4)\ \mathrm{nb^{-1}}$~\cite{ALICE:2018pqt}, 
and $1.49 \times 10^9$ minimum bias events taken at \thirteen between 2016 and 2018, corresponding to an integrated luminosity of $\mathcal{L}_{int}=(25.81\pm0.43)\ \mathrm{nb^{-1}}$~\cite{ALICE-PUBLIC-2021-005}.

The Monte Carlo samples used for the corrections, described in Section~\ref{Sec:corr}, were produced with the PYTHIA 6.4.25 event generator~\cite{Sjostrand:2006za}, with the Perugia 2011 tune~\cite{Skands:2010ak} and the GEANT 3.21.11~\cite{Brun:118715} transport model.
The ALICE detector layout and the variations of the data-taking conditions during the run were reproduced in the simulation. They shall be referred to as PYTHIA 6 and GEANT 3 in the following, unless otherwise specified. The reconstruction procedure of jets containing \Dzero mesons is briefly illustrated in the following section.

\section{\texorpdfstring{\Dzero}-meson tagged jet reconstruction and corrections}
\label{Sec:analysis}

\subsection{\texorpdfstring{\Dzero}-meson and jet reconstruction}

The \Dzero mesons were reconstructed via their hadronic decay channel \Dzero$\rightarrow$\kam\pip~\label{eqn:DzeroDecayChannel} and its charge conjugate $(\text{BR}=3.950\pm0.031\%$)~\cite{Zyla:2020zbs}.
The \Dzero meson and its anti-particle are treated equivalently and shall both be referred to as \Dzero in the following, unless otherwise specified. The \Dzero mesons produced directly in the charm-quark fragmentation or in decays of directly-produced excited charm hadron states are called prompt \Dzero mesons, and those that originate from decays of beauty hadrons are denoted as non-prompt \Dzero mesons. 

The \Dzero candidates were constructed by combining oppositely charged tracks identified as \pion or \kaon mesons.  
These tracks were required to have $p_\mathrm{T} > 300$ \MeVc, $|\eta|<0.8$, 
a minimum of 70 crossed rows in the TPC, 
with at least 80\% of these having an associated cluster of charged signals in the TPC end plates, and at least two hits in the ITS, with a minimum of one of these in the two innermost layers. 
For tracks with $p_\mathrm{T} < 3$ \GeVc, 
a hit in the innermost layer of the ITS was also required. 
With the mentioned kinematic selections on the pion and kaon tracks, 
the \Dzero-meson acceptance in rapidity 
is \ptd-dependent with the upper limit growing from $|y_{\text{D}}| = 0.5$ at $\ptd=0$ to $|y_{\text{D}}| = 0.8$ at $\ptd = 5$~\GeVc. 
The particle identification was carried out by exploiting the specific energy loss $\mathrm{d}E/\mathrm{d}x$ in the TPC and the time-of-flight provided by the TOF detector.
Pions and kaons were selected within $3\sigma$ (with $\sigma$ being the resolution on the $\mathrm{d}E/\mathrm{d}x$ and the time-of-flight) from the expected mean values. 
Particles with no TOF information were identified using the TPC information only. 
In order to reduce the combinatorial background, geometrical selections on the \Dzero-decay topology were applied, 
exploiting the displacement (typically of a few hundred $\upmu$m) of the \Dzero-meson decay 
vertices from the primary vertex of the interaction.
The selection was tuned to provide a high \Dzero signal-to-background ratio. Further details about the track and \Dzero-candidate selections can be found in Ref.~\cite{ALICE:2019nxm,Acharya:2019mno,Acharya:2019icl}.

For the jet reconstruction, charged particles were used and were required to have \pt $>$ 150 \MeVc and $|\eta|<0.9$. 
The track selection criteria applied were less stringent than those applied to the \Dzero-daughter tracks in order to ensure a flat acceptance in $\eta$ and $\varphi$.
Jets were reconstructed with the \antikt clustering algorithm as implemented in the FastJet package~\cite{Cacciari:2011ma} with the resolution parameters $R = 0.2, 0.4, 0.6$, using the \pt recombination scheme.
To ensure that the whole jet was contained within the detector acceptance, jets were required to have their axes within the pseudorapidity range of $\left |\etajet \right |~<~0.9-R$.
At low momenta, \Dzero-decay products can be emitted at angles larger than the defined jet cone size. 
In order to ensure that the \pion and \kaon mesons from the \Dzero decay were assigned to the same jet, 
they were removed from the set of charged-particle tracks before the jet reconstruction and their four-momenta were replaced by that of the \Dzero candidate.
A charm jet was tagged by the presence of a \Dzero-meson candidate among its constituents.  
In the rare case in which more than one \Dzero-meson candidate was present, the procedure was repeated separately for each \Dzero-meson candidate in the event.
No correction for the background coming from the underlying event was applied. 
The analysis procedure closely followed previous ALICE studies of charm jets tagged with \Dzero mesons~\cite{Acharya:2019zup}.

\subsection{Raw yield extraction}
\label{Sec:rawYields}

Raw yields of \Dzero jets were obtained using a statistical approach.
Oppositely charged kaons and pions from the decays of the \Dzero-meson candidates were combined and the pair's invariant mass distribution ($M$) was extracted in several intervals of \Dzero-meson transverse momentum within $2< \ptd <36$~\GeVc. 
For the \zch studies, the \Dzero-jet signal was also split in different ranges of \ptchjet.
The $M$ distributions were fitted with a function composed of a Gaussian for the \Dzero-signal peak and an exponential for the background. 
When two oppositely charged pion and kaon tracks are combined to form a \Dzero candidate, it may happen that neither the kaon nor the pion hypothesis can be definitively excluded for either of the tracks. 
In that case, the pair was accepted both as a \Dzero and a \Dzerobar candidate, and the two related invariant mass values, resulting from swapping the pion and kaon mass hypotheses for the two tracks, were considered in the analysis.
The candidates corresponding to a real \Dzero (or \Dzerobar) meson but with the wrong decay-product mass assignment are referred to as reflections.
The reflection component was included in the invariant mass fitting procedure and subtracted from the signal. 
The reflection templates were obtained from simulations with the PYTHIA~6 event generator and parametrised as a sum of two Gaussians with the means, widths, and the \Dzero signal-over-reflection ratio fixed to values obtained in the simulations.

\begin{figure}[tb]
    \begin{center}
    \includegraphics[width = \textwidth]{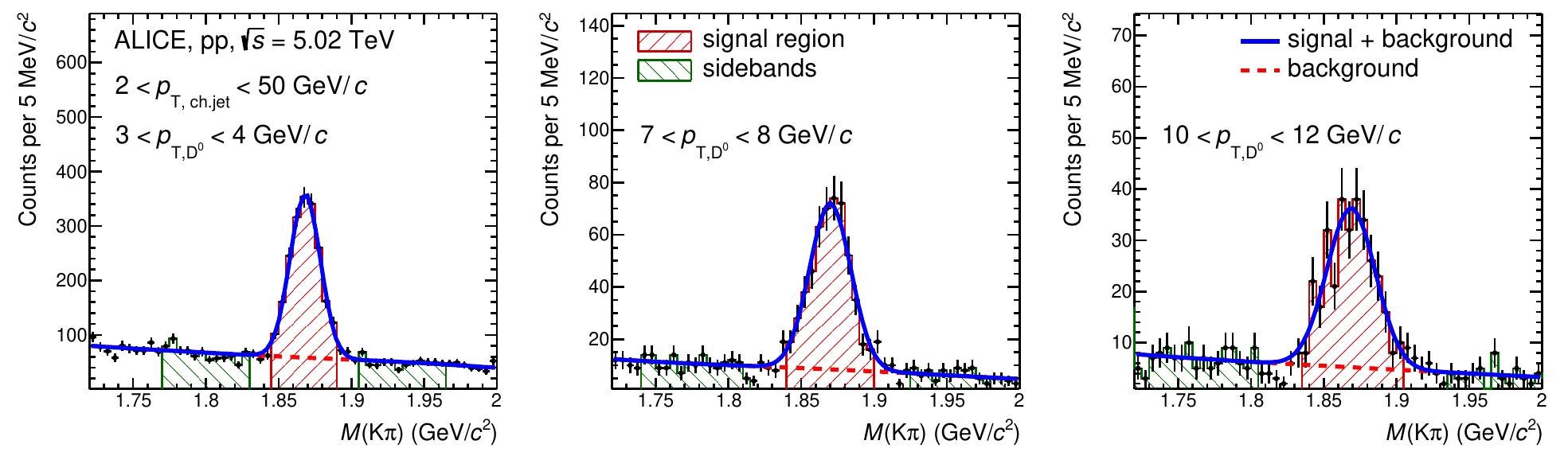}
    \includegraphics[width = \textwidth]{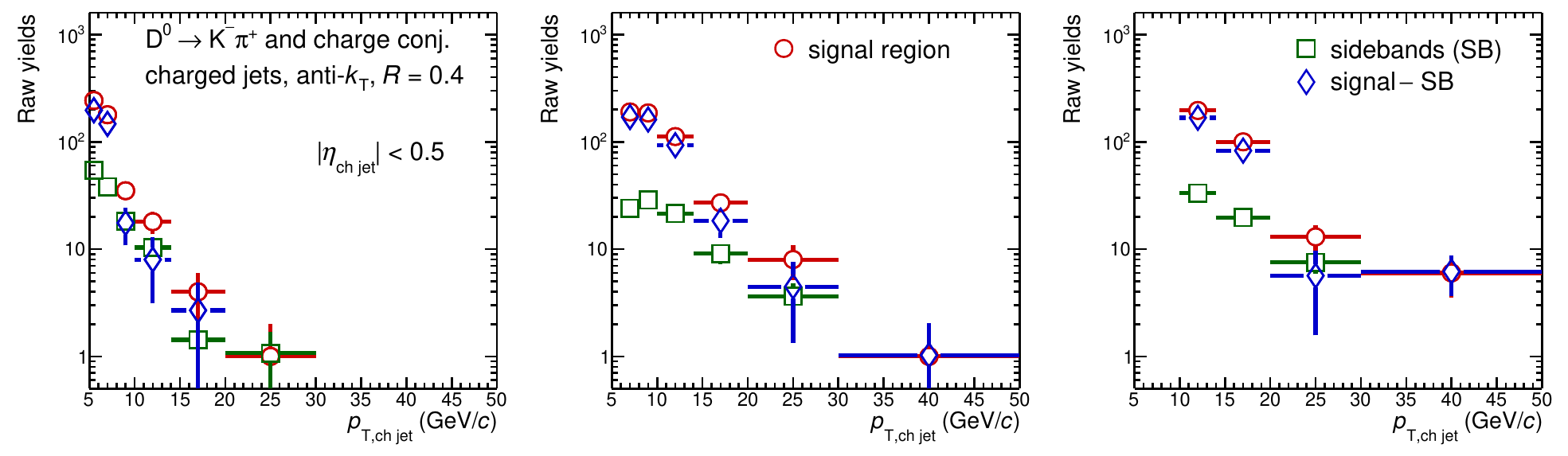}
    \end{center}
    \caption{Top: invariant mass distributions of \Dzero-jet candidates for $2<$\ptchjet$<50$ \GeVc and $R=0.4$ in \pp collisions at \five, in the \Dzero-meson transverse momentum intervals: $3<$\ptd$<4$~\GeVc (left), $7<$\ptd$<8$~\GeVc (centre), and $10<$\ptd$<12$~\GeVc (right). 
    The total fit function is represented by the blue solid line, while the red dashed line represents the sum of the background and reflection fit functions. The red and green shaded areas correspond to the peak and sideband regions, respectively. Bottom: \Dzero-jet raw yields as a function of \ptchjet in the signal and sideband regions, and their subtracted yields.}
    \label{fig:rawYieldsJet}
\end{figure}

\begin{figure}[tb]
    \begin{center}
    \includegraphics[width = \textwidth]{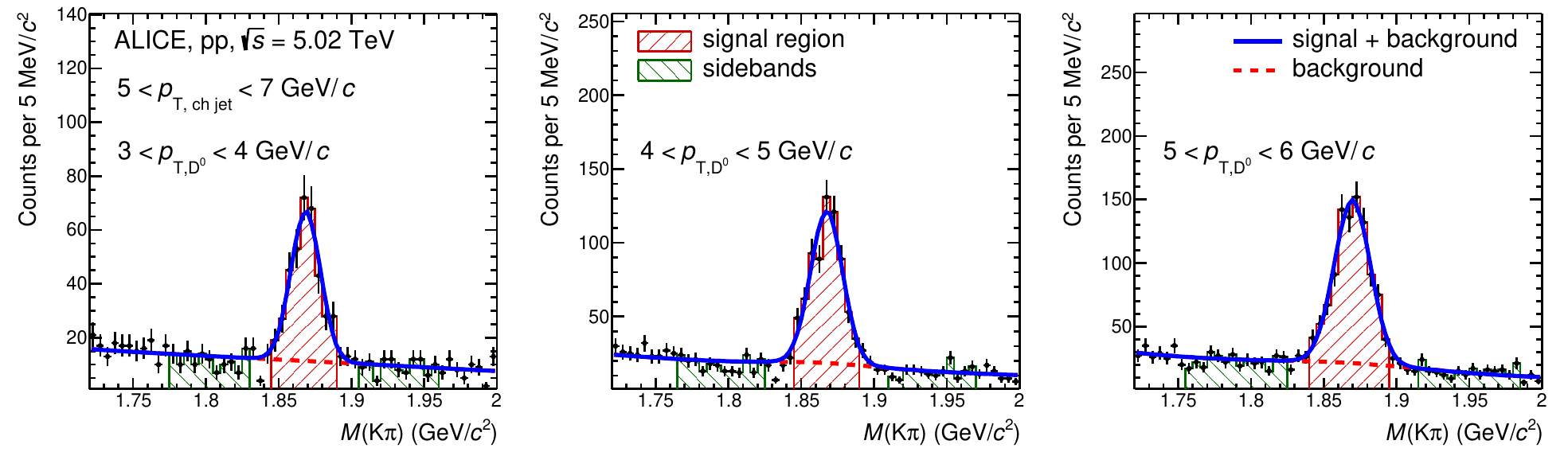}
    \includegraphics[width = \textwidth]{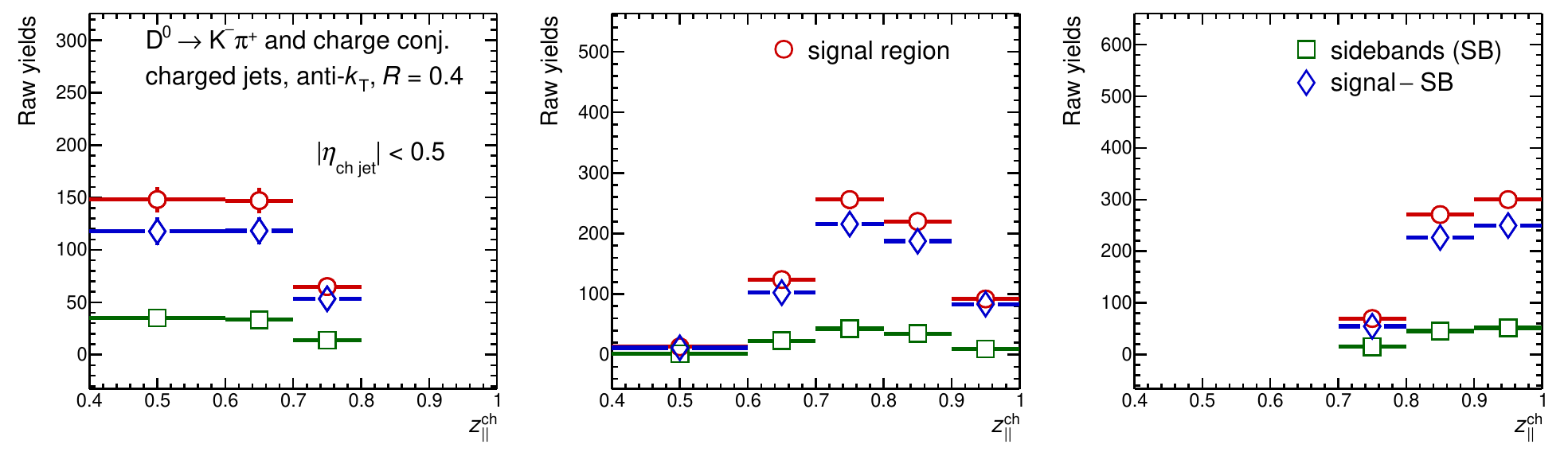}
    \end{center}
    \caption{Top: invariant mass distribution of \Dzero-jet candidates for one jet-\pt interval of ${5<\ptchjet<7~\GeVc}$ and $R=0.4$ in \pp collisions at \five, in \Dzero-meson transverse momentum intervals: $3<\ptd<4~\GeVc$ (left), $4<\ptd<5$~\GeVc (centre), and $5<\ptd<6$~\GeVc (right). The total fit function is represented by the blue solid line, while the red dashed line represents the background fit function. The red and green shaded areas correspond to the peak and sideband regions, respectively. Bottom: \Dzero-jet raw yields as a function of \zch in the signal and sideband regions, and their subtracted yields.}
    \label{fig:rawYieldsZ}
\end{figure}

The signal region was defined to be within $\left | M - \mu_{\textrm{fit}} \right |<2 \sigma_{\textrm{fit}}$, where $\mu_{\textrm{fit}}$ is the mean and $\sigma_{\textrm{fit}}$ is the width of the Gaussian fit component, respectively. 
The background regions (sidebands) were chosen as follows: $4\sigma_{\textrm{fit}}<|M - \mu_{\textrm{fit}} |<9\sigma_{\textrm{fit}}$. 
The top panels of Fig.~\ref{fig:rawYieldsJet} and~\ref{fig:rawYieldsZ} show examples of $M$ distributions for different intervals of \ptd. 
The signal and sideband regions are represented by the dashed red and green areas, respectively. The reflection contributions are included in the background fit function.
The bottom panels of Fig.~\ref{fig:rawYieldsJet} and~\ref{fig:rawYieldsZ} present the raw yields of \Dzero jets as a function of \ptchjet and \zch extracted for the signal and sideband $M$ regions in each \ptd interval. 
The sideband distributions were normalised to the background yield in the peak region and subtracted from the signal-region distributions in order to obtain the raw \mbox{\Dzero-jet} \ptchjet and \zch distributions.

\subsection{Corrections}
\label{Sec:corr}

A threefold correction was applied to the raw \Dzero-jet \ptchjet and \zch distributions. 
The corrections account for: (i) the efficiency and acceptance of the \Dzero-jet reconstruction, (ii) the contribution of \Dzero mesons originating from b-hadron decays, and (iii) the momentum smearing introduced by detector effects. 
The systematic uncertainties of these corrections are discussed in Sec.~\ref{Sec:sys}.

\subsubsection{Reconstruction efficiency}

\begin{figure}[tb]
     \centering
     \begin{minipage}[tbh]{0.49\textwidth}
         \centering
         \includegraphics[width=\textwidth]{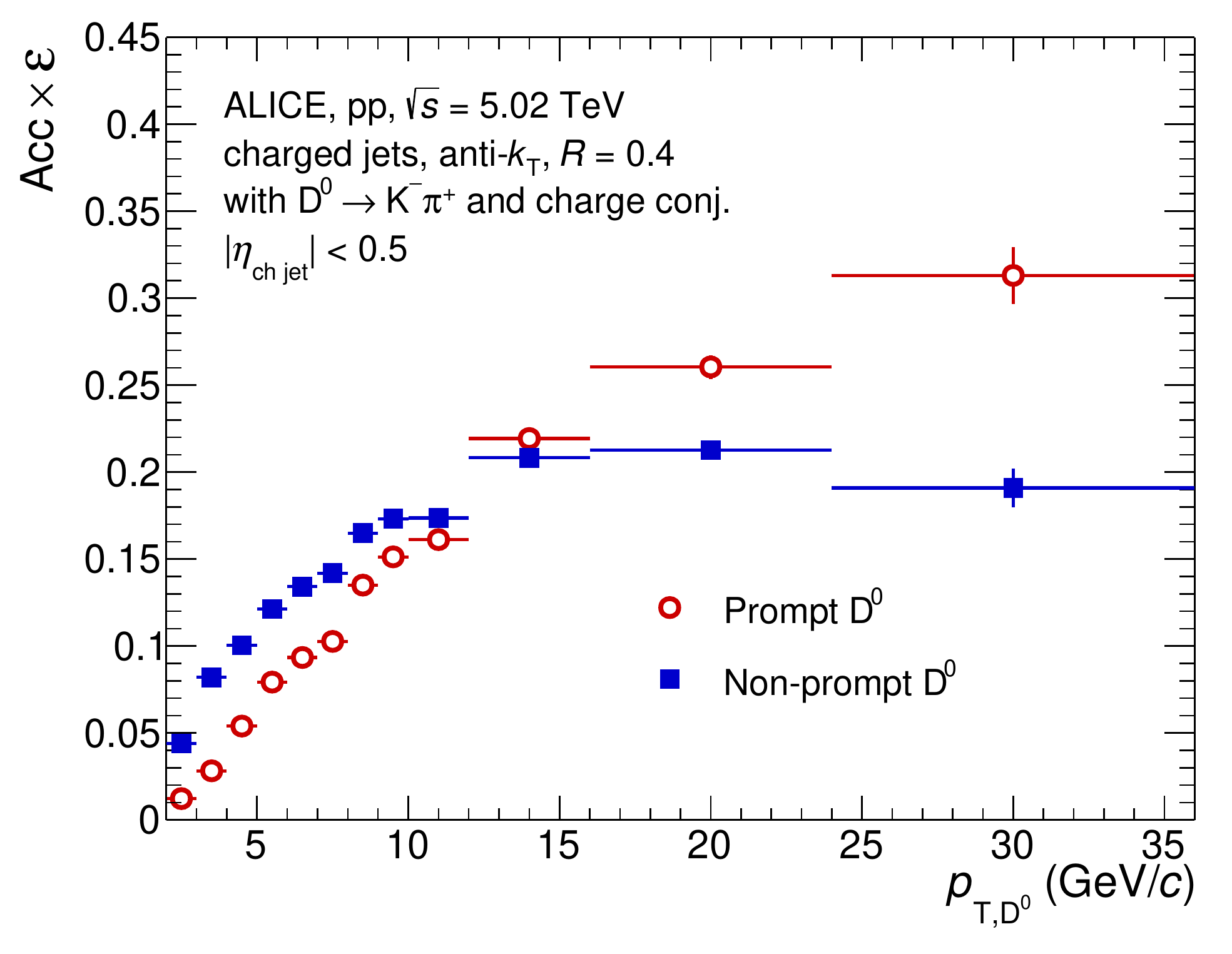}
     \end{minipage}
     \hfill
     \begin{minipage}[tbh]{0.49\textwidth}
         \centering
         \includegraphics[width=\textwidth]{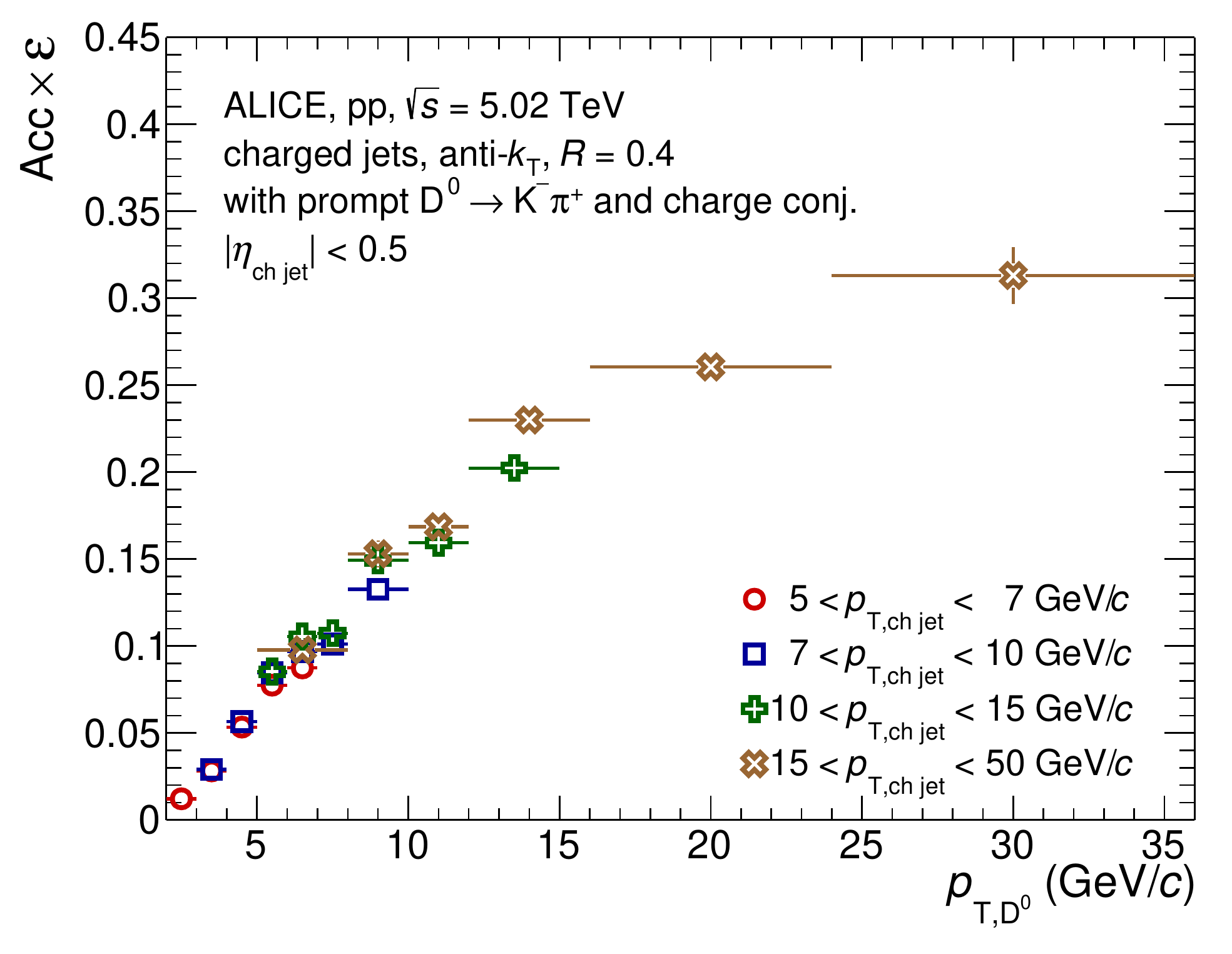}
     \end{minipage}
     \caption{Product of acceptance and efficiency, $\textrm{Acc}\times\epsilon$, for \Dzero-jet reconstruction as a function of \ptd with $R=0.4$ in \pp collisions at \five. Left: $\textrm{Acc}\times\epsilon$ for prompt and non-prompt \Dzero jets in the range \mbox{$5<$~\ptchjet~$<50$~\GeVc.} Right: $\textrm{Acc}\times\epsilon$ for prompt \Dzero jets in different jet-\pt intervals.
     }
     \label{fig:DjetEfficiencyJetandZ}
\end{figure}

The reconstruction efficiency of the \Dzero jets within the detector acceptance was calculated using the simulation described in Section~\ref{Sec:detectoranddata}. The efficiency was defined as the ratio of \Dzero jets that passed all the data analysis selection requirements to all generated \Dzero jets within $|\etajet| < 0.9-R$.
The efficiency depends on the \Dzero-meson topological selections, which are stricter at low \ptd in order to reduce the larger combinatorial background present in this kinematic region. Therefore, the \Dzero-jet reconstruction efficiency depends strongly on \ptd, but has negligible dependence on \ptchjet in the measured ranges. 
Fig.~\ref{fig:DjetEfficiencyJetandZ} (left) shows the product of acceptance and efficiency for prompt and non-prompt \Dzero jets. 
The acceptance and efficiency for non-prompt \Dzero jets tends to be higher than that for prompt \Dzero jets at low \ptd with a crossing point around $\ptd=15 $ \GeVc. 
The non-prompt \Dzero mesons are selected with higher efficiency because of their larger displacement from the primary vertex. 
However, at higher \ptd, a selection on the impact parameters of the decay particles suppresses the non-prompt contribution while keeping most of the prompt ones. 
Both efficiencies are independent of the \ptchjet selection as is seen for the prompt efficiencies in different analysed \ptchjet intervals in Fig.~\ref{fig:DjetEfficiencyJetandZ} (right).

The product of the acceptance (Acc) and the reconstruction efficiency ($\epsilon$) of the prompt \Dzero jets was used to correct the raw yields extracted in different intervals of \ptd, as described in Section~\ref{Sec:rawYields}. 
The efficiency-corrected \ptchjet distributions were then summed over all the \ptd intervals, according to
    
\begin{equation}
  N(\ptchjet) = \sum_{\ptd} \frac{ N_{\textrm {raw}}(\ptchjet, \ptd)}{(\textrm{Acc}\times\epsilon)_{\textrm{c}}(\ptd)},
\label{eq:EffCorr}
\end{equation}
where c represents charm (prompt \Dzero mesons) and $N$ is the total efficiency-corrected yield.
A similar method was also used to extract efficiency-corrected \zch distributions in different \ptd and \ptchjet intervals.

\subsubsection{Subtraction of b-jet contribution}\label{subsec:bjet}
Since the natural fraction of \Dzero mesons originating from b-quark fragmentation is biased by the applied topological selection criteria, 
the non-prompt \Dzero-meson contribution was subtracted from the reported distributions to get the desired prompt \Dzero-jet distributions ($N^{\textrm{c}}$).
The limited sample size did not allow for a data-driven estimation of the non-prompt \Dzero-jet fraction. Therefore, NLO pQCD calculations of POWHEG~\cite{Nason:2004rx,Frixione:2007vw,Alioli:2010xd,Alioli:2010xa} coupled to the PYTHIA~6~\cite{Sjostrand:2006za,Skands:2010ak} MC parton shower were used to estimate this contribution. 
There are three parameters in the calculations: the beauty-quark mass ($m_{\textrm {b}}$) that was set to $m_{\textrm {b}}=4.75$~\GeVcsq, and the renormalisation ($\mu_{\textrm {R}}$) and factorisation  ($\mu_{\textrm {F}}$) scales, both set to the quark transverse mass $\mu_{\textrm {R}}=\mu_{\textrm {F}}=\sqrt{m_{\textrm {b}}^2+p_{\textrm {T}}^2}$. 
Parton distribution functions (PDF) obtained from the CT10NLO set~\cite{Lai:2010vv} using the LHAPDF6~\cite{Buckley:2014ana} interpolator were used.

\begin{figure}[tb]
     \centering
     \begin{minipage}[tbh]{0.49\textwidth}
         \centering
         \includegraphics[width=\textwidth]{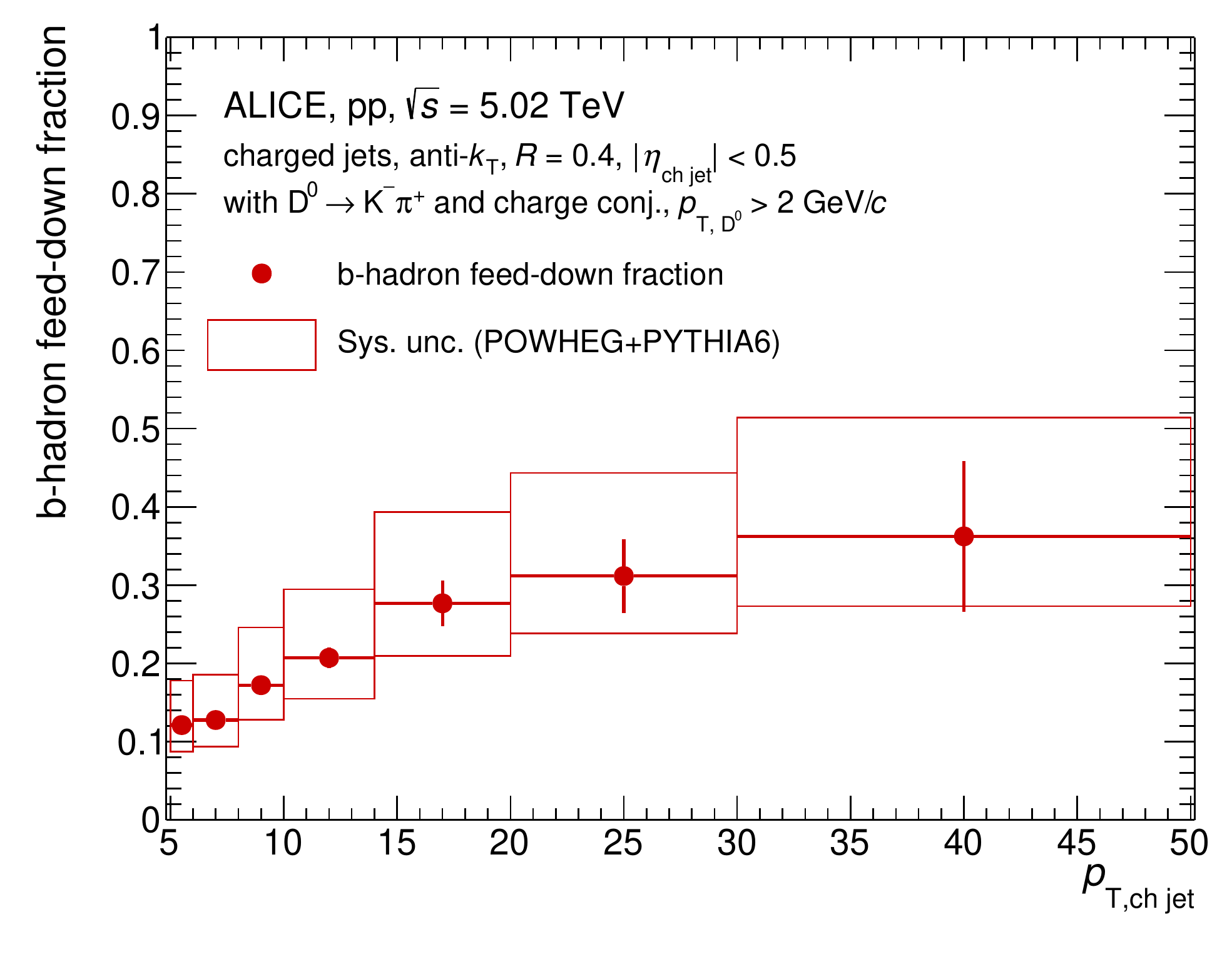}
     \end{minipage}
     \hfill
     \begin{minipage}[tbh]{0.49\textwidth}
         \centering
         \includegraphics[width=\textwidth]{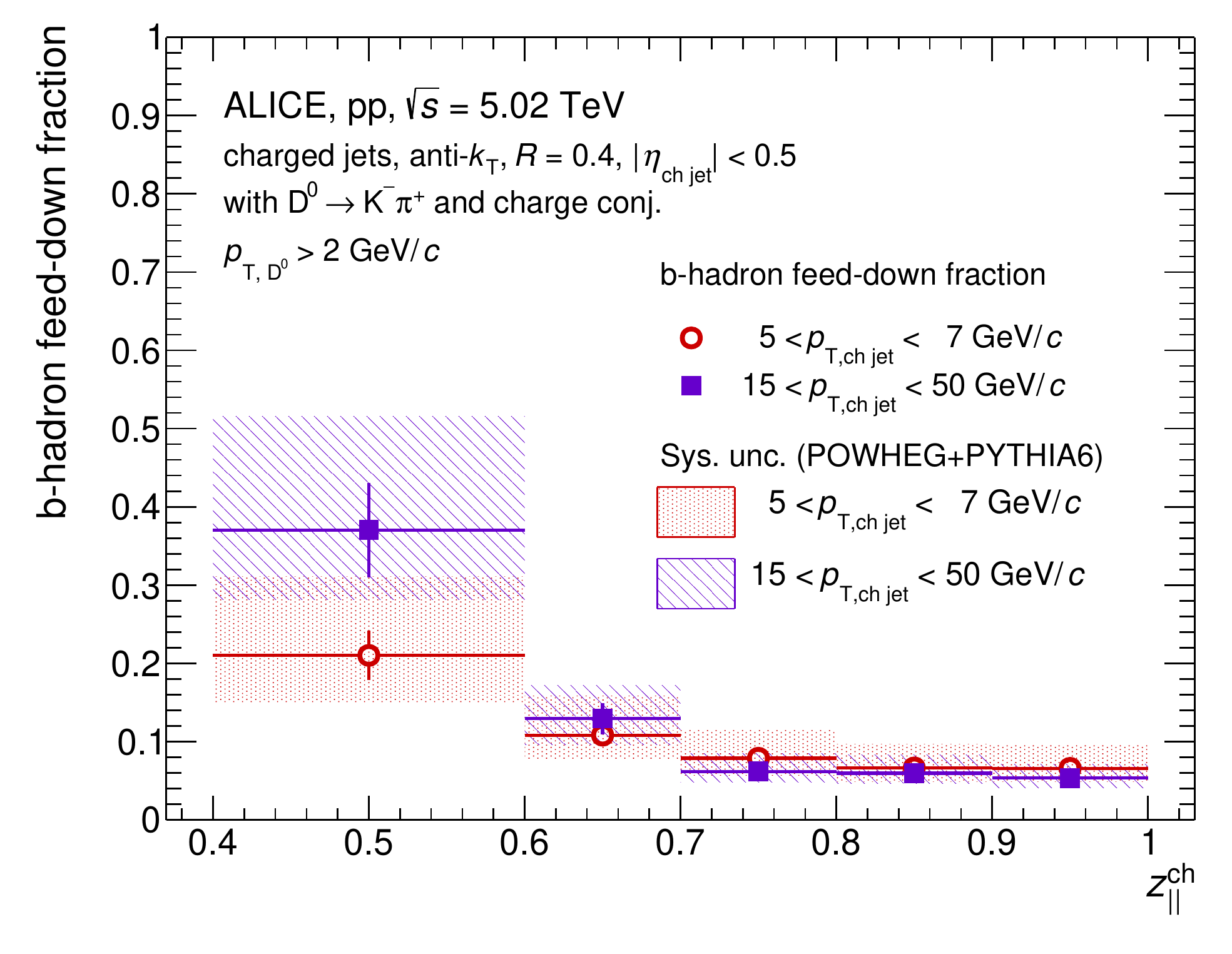}
     \end{minipage}
     \caption{Feed-down fraction of \Dzero jets from b-hadrons in \pp collisions at \five for $R=0.4$ as a function of \ptchjet in $5< \ptchjet <50$~\GeVc (left) and as a function of \zch in two \ptchjet intervals $5< \ptchjet <7$~\GeVc and $15< \ptchjet <50$~\GeVc (right).}
     \label{fig:FDfraction}
\end{figure}

The simulation output was scaled by the ratio of the reconstruction efficiencies of non-prompt and prompt \Dzero jets ($\epsilon_{\textrm{b}}/\epsilon_{\textrm{c}}$) 
because $\epsilon_{\textrm{b}}$-scaled non-prompt simulations ($\epsilon_{\textrm{b}}\times N^{\textrm{b}}_{\textrm{POWHEG}}$)  are comparable with $\epsilon_{\textrm{c}}$-scaled prompt \Dzero-jet distributions ($\epsilon_{\textrm{c}}\times N^{\textrm{c}}$).
In the next step, the POWHEG + PYTHIA 6  \ptchjet (\zch) distributions were smeared using a response matrix (RM) for non-prompt \Dzero jets, $\text{RM}_{\textrm {b}\rightarrow\Dzero}$, which maps the \Dzero-jet particle-level variables (\ptchjetgen,~\zchgen) from PYTHIA~6 simulations to the detector-level variables (\ptchjetdet,~\zchdet) reconstructed in full PYTHIA~6 + GEANT~3 detector simulations. 
The RM was also re-weighted by the prompt \Dzero-jet efficiency to address the fact that the measured sample is already corrected by it.
A correction was made to account for jets which were inside the detector acceptance but outside the generated range, 
and for those which were outside of the acceptance but inside the generated range. 
The calculated b-jet feed-down fraction in the measured sample is shown in Fig.~\ref{fig:FDfraction} as a function of \ptchjet and \zch. 
The estimation of the corresponding systematic uncertainties shown in Fig.~\ref{fig:FDfraction} is described in Section~\ref{Sec:sys}.

The b-hadron feed-down contribution was then subtracted from the efficiency-corrected \ptchjet  distributions according to

\begin{eqnarray}
N^{\textrm{c}}(\ptchjetdet) 
=&& N(\ptchjetdet) -\sum_{\ptd}
\text{RM}_{\textrm {b}\rightarrow\Dzero}(\ptchjetdet,\ptchjetgen,\ptd) \nonumber\\
&&\otimes \sum_{\ptd} \frac{(\textrm{Acc}\times\epsilon)_{\textrm{b}}(\ptd)}{(\textrm{Acc}\times\epsilon)_{\textrm{c}}(\ptd)} N^{\textrm{b}}_{\textrm{POWHEG}}(\ptd,\ptchjetgen),
\label{eq:Bsub}
\end{eqnarray}

where:
\begin{itemize}
    \item c and b stand for charm (prompt \Dzero) and beauty (non-prompt \Dzero), respectively;
	\item $N(\ptchjetdet)$ 
	is the total efficiency-corrected measured yield, before subtraction of the b-jet contribution;
	\item $N^{\textrm{c}}(\ptchjetdet)$
	is the efficiency-corrected measured yield after subtraction of the b-jet contribution; 
	\item the symbol $\otimes$ should be interpreted as the convolution of the non-prompt $\text{RM}_{\textrm {b}\rightarrow\Dzero}$ and the vector of the yields;
	\item $(\textrm{Acc}\times\epsilon)_{\textrm{c}}(\ptd)$, $(\textrm{Acc}\times\epsilon)_{\textrm{b}}(\ptd)$ are the \ptd-dependent products of the acceptance and the reconstruction efficiency for prompt and non-prompt \Dzero jets respectively;
	\item $N^{\textrm{b}}_{\textrm {POWHEG}}(\ptd,\ptchjetgen)$ is the non-prompt \Dzero-jet \ptchjet cross section from the POWHEG simulation scaled by the integrated luminosity of the analysed data.
\end{itemize}
An analogous subtraction was also performed for the \zch studies.

\subsubsection{Unfolding}
\label{subsec:unfold}

\begin{figure}[tb]
     \centering
     \begin{minipage}[tbh]{0.45\textwidth}
         \centering
         \includegraphics[width=\textwidth]{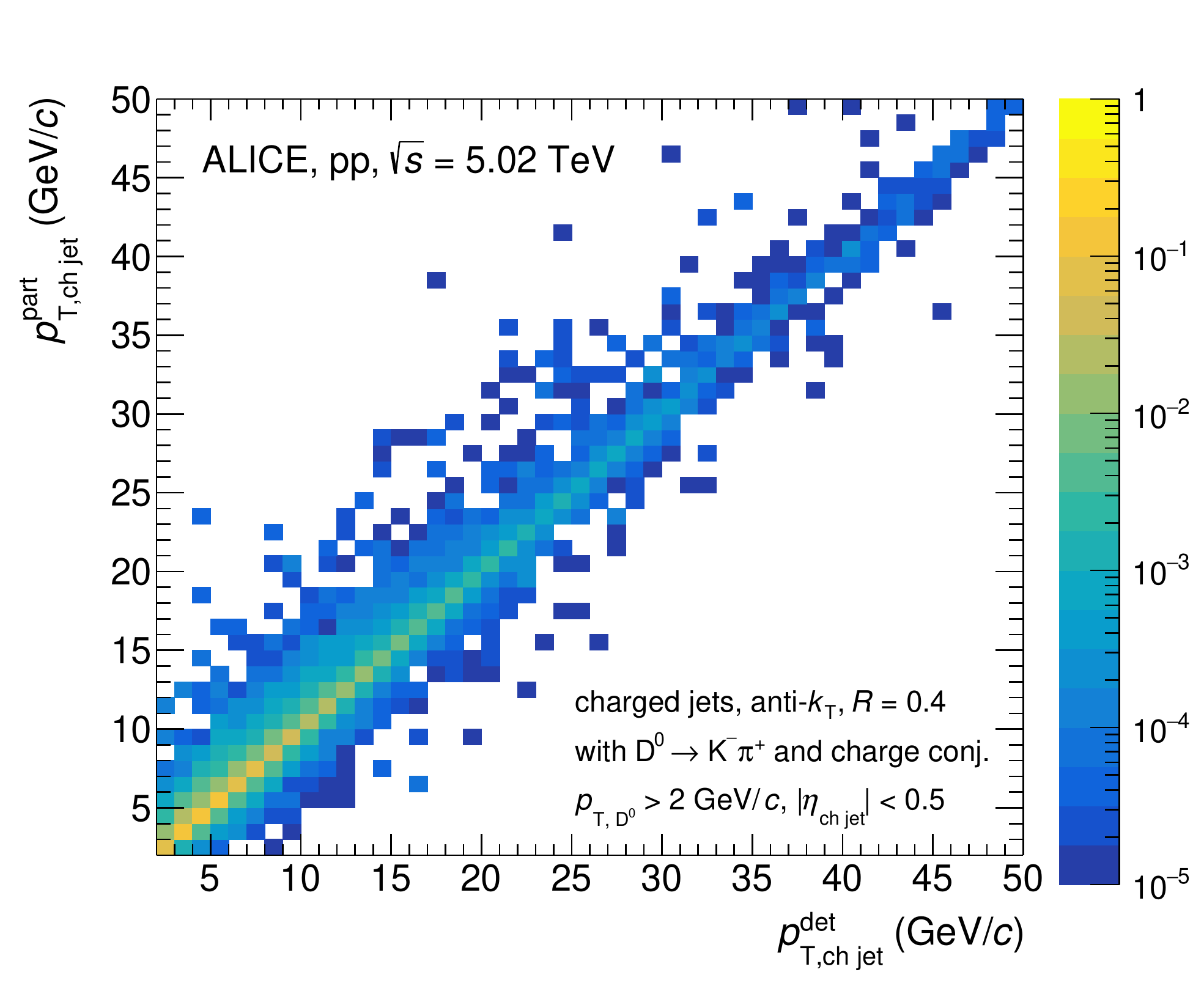}
     \end{minipage}
     \hfill
     \begin{minipage}[tbh]{0.54\textwidth}
         \centering
         \includegraphics[width=\textwidth]{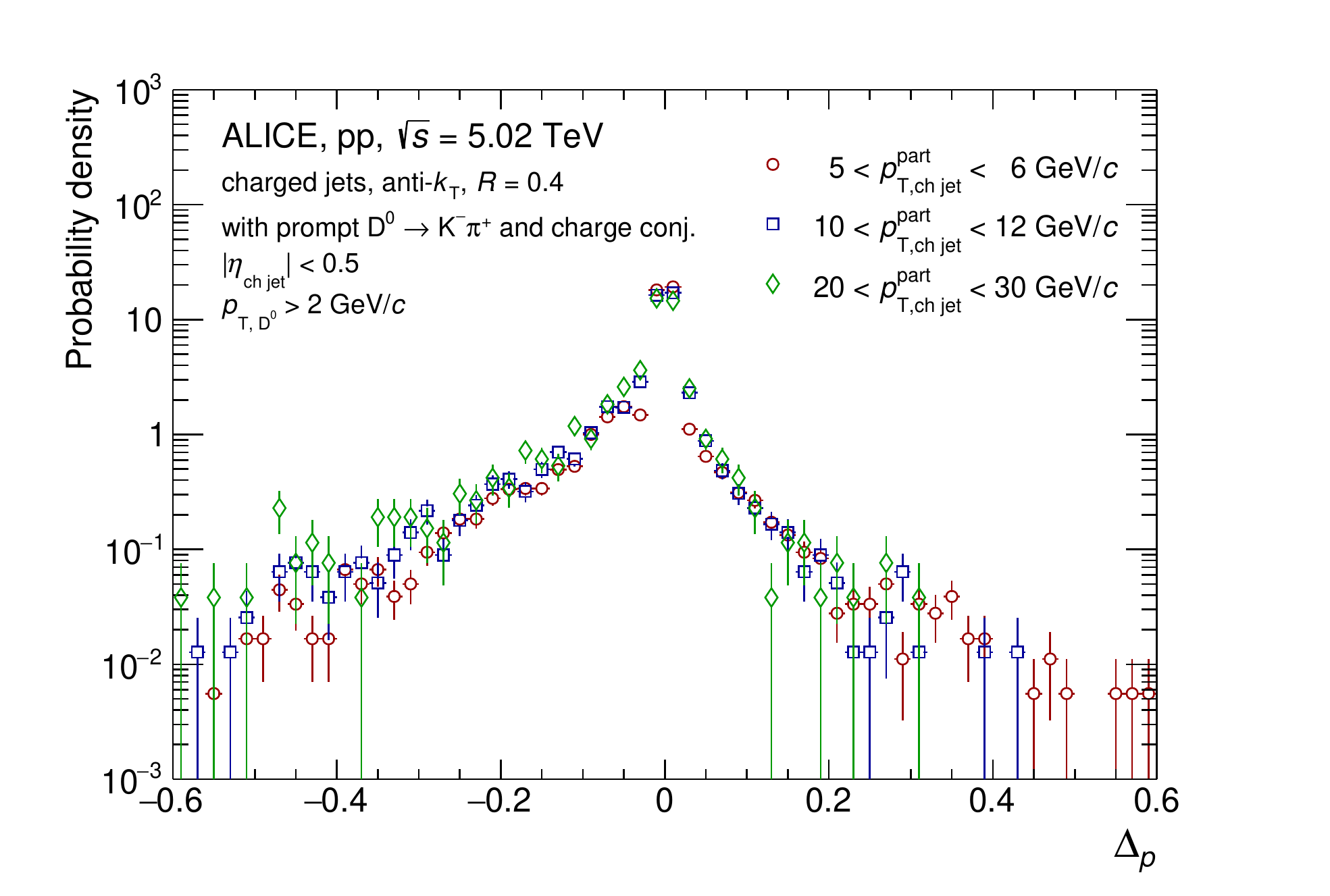}
     \end{minipage}
     \caption{Left: detector response matrix of matched jets used for unfolding the \ptchjet distribution with $R=0.4$ in \pp collisions at \five in the range $2< \ptchjet <50$ \GeVc.
     Right: \ptchjet resolution, $\Delta_{p}$, 
     for $5< \ptchjetgen <6$~\GeVc, $10< \ptchjetgen <12$~\GeVc, and $20< \ptchjetgen <30$~\GeVc.}
     \label{fig:RMjet}
\end{figure}

\begin{figure}[tb]
     \centering
     \begin{minipage}[tbh]{0.46\textwidth}
         \centering
         \includegraphics[width=\textwidth]{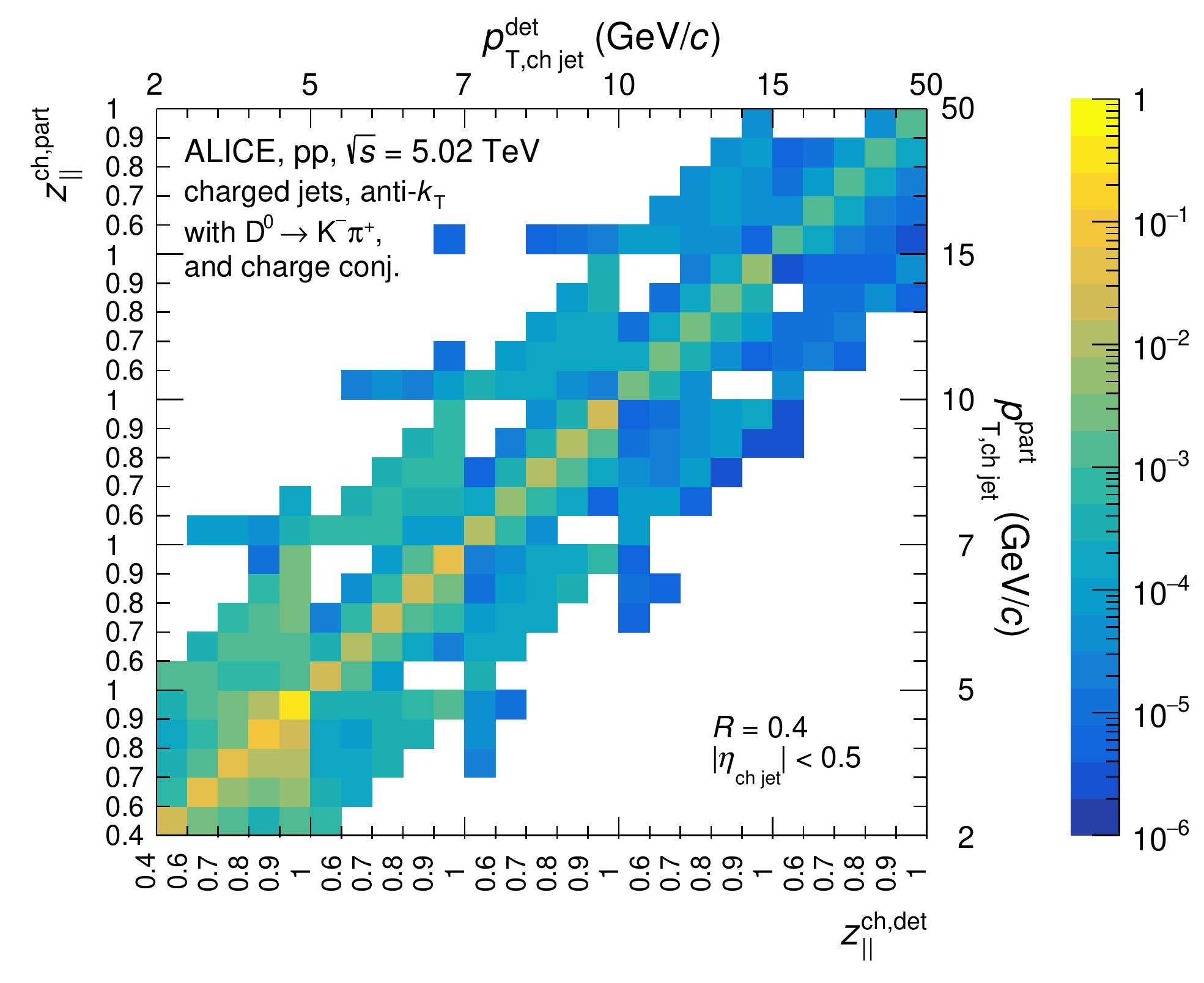}
     \end{minipage}
     \hfill
     \begin{minipage}[tbh]{0.53\textwidth}
         \centering
         \includegraphics[width=\textwidth]{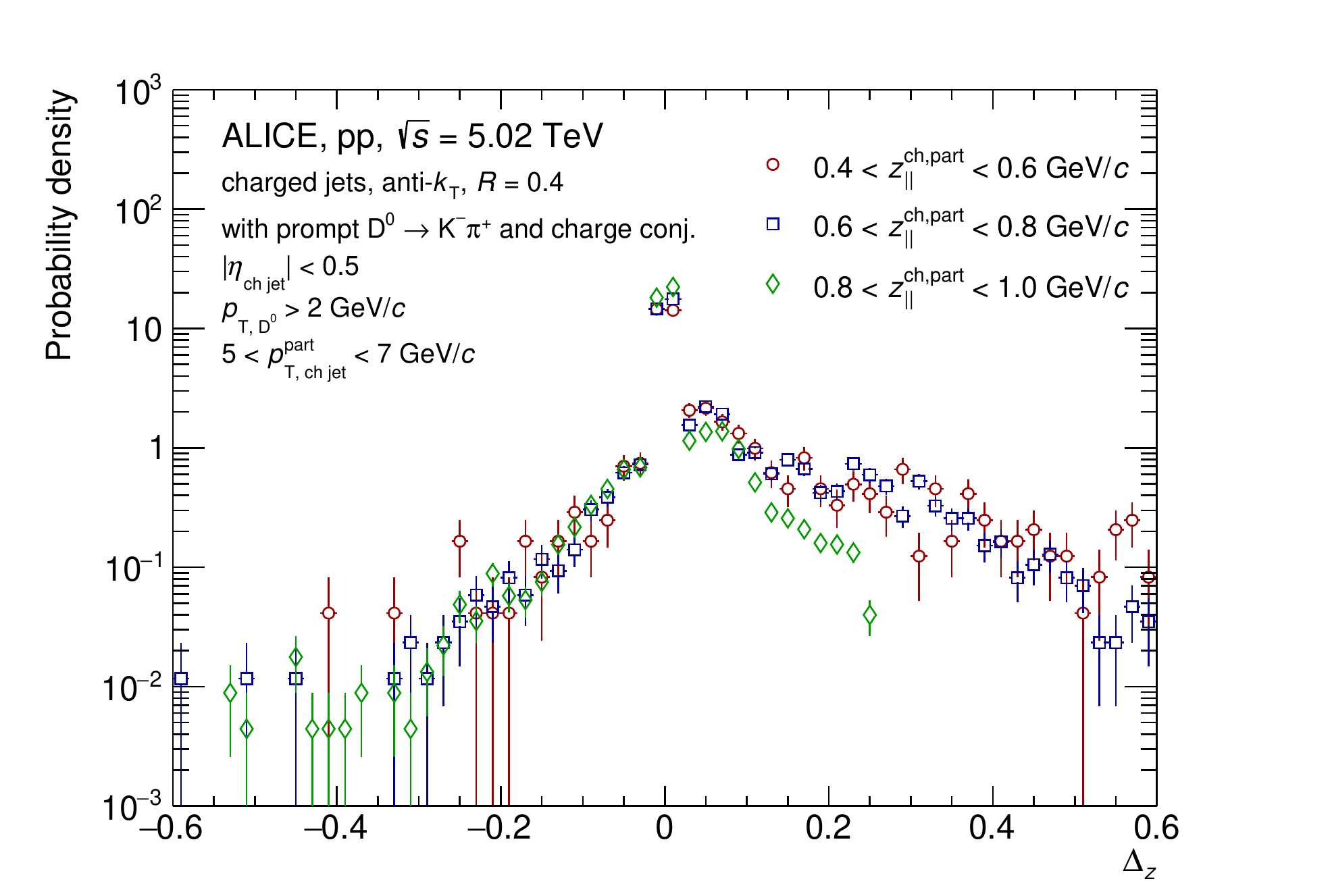}
     \end{minipage}
     
     \caption{Left: detector response matrix of matched jets used for unfolding \zch distribution with $R=0.4$ in \pp collisions at \five in the range $2< \ptchjet <50$ \GeVc and $0.4< \zch <1$. Right: \zch resolution, $\Delta_{z}$, in the \zchgen intervals $0.4$--$0.6$, $0.6$--$0.8$, and $0.8$--$1$ for the \ptchjet interval $5< \ptchjetgen <7$~\GeVc.
     }
     \label{fig:RMz}
\end{figure}

The measured \ptchjet and \zch distributions were corrected for the detector resolution and track momentum smearing. 
The corrections were encoded in a detector RM that mapped the \Dzero-jet particle-level variables (\ptchjetgen,~\zchgen) from PY\-THIA~6 simulations to the detector-level variables (\ptchjetdet,~\zchdet) reconstructed in full PYTHIA~6 + GEANT~3 detector simulations. The detector and particle-level charged-particle jets were matched by requiring the same prompt \Dzero meson among their constituents. The jets at both levels were reconstructed using the \antikt clustering algorithm.
The 1-dimensional RM used to correct for the \ptchjet and the corresponding relative resolution, defined as ${\Delta_p=(\ptchjetdet-\ptchjetgen)/\ptchjetgen}$,
are displayed in Fig.~\ref{fig:RMjet}. 
Fig.~\ref{fig:RMz} presents the 2-dimensional RM for the \zch and \ptchjet variables, along with the relative resolution ${\Delta_z=(\zchdet-\zchgen)/\zchgen}$ for a given \ptchjet interval.

The finite detector resolution modifies the measured yields as a function of \ptchjet and \zch. They were therefore unfolded using an iterative method based on Bayes' theorem~\cite{dagostini2010improved} as implemented in the RooUnfold package~\cite{RooUnfold}. The \ptchjet spectra were unfolded using a 1D unfolding method, while for the \zch distributions a 2D method was implemented. The RM was scaled by the prompt \Dzero-jet efficiency before unfolding the measured spectra. Five iterations showed to be optimal, representing a good convergence of the unfolding, and were chosen as the default.
The unfolding was performed in the following ranges: 2~$<\ptchjet<$~50~\GeVc and 0.4~$<\zch<$~1. 
Similar to the correction of non-prompt \Dzero-jet simulations, the measured spectra were also corrected for in order to account for jets which were inside the detector acceptance but outside the generated range, and for those which were outside of the acceptance but inside the generated range. 
This resulted in a correction of about 1--2\%.

To verify the stability of the unfolding and the choice of the number of iterations, several checks were performed. 
Firstly, the unfolded spectra were folded back and compared to the original data. 
A good agreement was found in all the cases. 
Secondly, Pearson correlation coefficients were calculated to determine the optimal number of iterations and lastly, a closure test was performed which also provided an estimate of the systematic uncertainty of the method and is described in more detail in Section~\ref{Sec:sys}. 
While the reported range is 5~$< \ptchjet <$~50~\GeVc, the measurements in the 2~$< \ptchjet< $~5~\GeVc interval were kept in the unfolding for both \ptchjet and \zch\, to avoid potential biases due to edge effects.

\section{Systematic uncertainties}
\label{Sec:sys}

Several sources of systematic uncertainties were studied and can be separated into the following groups:
    (i) \Dzero-meson selections,
    (ii) raw yield extraction,
    (iii) beauty feed-down,
    (iv) unfolding,
    (v) track-reconstruction efficiency, and
    (vi) normalisation.

Discrepancies between data and simulations for the distributions of variables used in the \Dzero-meson selections can impact on the \Dzero-jet reconstruction efficiency. 
In order to assign a systematic uncertainty from this source, the \Dzero-meson topological selections were varied and the whole analysis procedure was repeated for each variation. 
The test spanned a variation of the reconstruction efficiency between 10\% and 25\%, depending on the \Dzero-meson \pt. 
The uncertainty was estimated by taking the root-mean-square of the results obtained with the different \Dzero-meson selection criteria. 
The uncertainty increases with \ptchjet and decreases with \zch, and varies between 1\% and 10\% for the sample at \thirteen and between 3\% and 25\% for \five.
The particle identification related systematic uncertainties for \Dzero-meson selections were negligible~\cite{ALICE:2019nxm} and excluded from the calculation.
    
The stability of the raw yield extraction procedure described in Section~\ref{Sec:rawYields} was assessed by performing multiple trials of the invariant mass fit while varying the fitting conditions. The conditions that were varied are: (i) the assumed shape of the background function (default exponential was replaced by linear and polynomial functions), (ii) the fit ranges, and (iii) the width ($\sigma_{\textrm{fit}}$) and (iv) mean ($\mu_{\textrm{fit}}$) of the Gaussian signal, which were left as free parameters or fixed to the MC values. 
The yields obtained from the multiple trials were compared to the default one and the root-mean-square of the relative differences was taken
as a part of the systematic uncertainties from the raw yield extraction. 
Secondly, the signal range was varied between 2 to 3 standard deviations of the signal peak width, while 
the sideband extraction range $|M - \mu_{\textrm{fit}} |$ was varied through 
4--9, 4--8, 3.5--9, 3.5--8, 4.5--9, and 4.5--8 units of standard deviation.
The corresponding uncertainties amount to about 1\% and 2\%, respectively. A third contribution to the systematic uncertainty on the raw yield extraction was assigned by varying the relative contribution of reflections by $\pm50$\% and the maximum deviation in the raw yield was taken as a systematic uncertainty.
The total uncertainty on the raw yield extraction was estimated to be $2$--$9$\% for the \ptchjet-differential cross section and $2$--$6$\% for the \zch distributions for the \thirteen analysis. 
For the \five analysis, the uncertainties reach a maximum of about 13\% for the \ptchjet-differential cross section, and are within 8\% for the \zch distributions.
  
The systematic uncertainty from the subtraction of the b-hadron decay contribution was determined by varying the parameters of the non-prompt \Dzero-jet POWHEG~+~PYTHIA~6 simulations.
They were varied individually in the following ways: (i) the beauty-quark mass was changed to 4.5~\GeVcsq and 5~\GeVcsq from the default 4.75~\GeVcsq and (ii) $\mu_{R}$ and $\mu_{F}$ were either halved or doubled from their nominal values, which were defined as the transverse mass of the beauty quark. 
The largest upward and downward variations of the resulting cross sections were taken as the systematic uncertainties. 
The uncertainty on the prompt cross section due to the feed-down subtraction was estimated for the \ptchjet-differential cross section to be $5$--$30$\% for the sample at \thirteen and $4$--$40$\% for \five.
For the \zch distributions it was $2$--$20$\% and $2$--$15$\% for \thirteen and \five, respectively. 
    
The systematic uncertainty on the unfolding procedure was assigned based on a MC closure test. 
The MC sample was randomly split into two subsamples and one part was used to build the RM while the other one was used as a test sample. 
The efficiency correction was applied on the test sample following the method used in the data analysis. 
The same unfolding procedure was then applied as in the data analysis, and the resulting distributions were compared to the generator-level MC distributions. 
The random split was performed ten times and a mean value of the deviations from these trials was taken as the final uncertainty of the unfolding procedure. 
The resulting uncertainty is $1$--$5$\% in most cases while rising with increasing \ptchjet and falling with increasing \zch. 
Occasionally, the uncertainty goes above 10\% for the highest \ptchjet interval in \ptchjet distributions and for lowest \zch intervals in \zch distributions.
In addition, several checks were performed in order to test the stability of the unfolding procedure explained in Section~\ref{subsec:unfold}, and were treated as a procedure cross-check: 
(i) the default number of five iterations of the Bayesian unfolding was varied by $\pm$1, 
(ii) the default MC generator-level prior distribution shape was varied by using the measured \ptchjet distribution or different parametrised power-law functions, $f(\ptchjet) = \ptchjet^{-a}e^{-ab/\ptchjet}$ with $3<a,b<5$, and (iii) the true and measured ranges for \ptchjet spectra provided to the unfolding procedure were varied.
All these tests gave consistent unfolding results with maximum relative deviations of 1\%.

The measurement is also affected by uncertainties on the efficiency of the track reconstruction that influence the jet momentum resolution and the \Dzero-meson reconstruction efficiency. 
The relative uncertainty on the reconstruction efficiency for a single track used for the jet reconstruction was estimated to be 4\%.
To assess the systematic uncertainty on the prompt cross section due to this source, a new detector RM was built where 4\% of all the reconstructed charged tracks in the detector simulations were randomly rejected.
The \ptchjet and \zch distributions were then unfolded using this modified RM and the results were compared to the final distributions unfolded with the default detector RM. The relative uncertainty from this source was found to increase with \ptchjet reaching a maximum of 10\%.
The uncertainties originating from the track momentum resolution were previously studied and found to be negligible~\cite{Acharya:2019tku, Abelev:2014ffa}. 
For the reconstruction efficiency of \Dzero-mesons, a \ptd-independent systematic uncertainty of 5\% was assigned based on the \Dzero-meson studies reported in Ref.~\cite{ALICE:2019nxm}.
Since the reported \zch distributions are self-normalised, this uncertainty is negligible in this case.

Finally, the normalisation of the \ptchjet-differential cross section was affected by a 0.8\% uncertainty on the \Dzero-meson decay branching ratio and by the uncertainty on the luminosity determination which is 2.1\% and 1.7\% for \five and \thirteen, respectively. 
  
The relative systematic uncertainties for \Dzero jets on their \ptchjet-differential cross sections for $R=0.4$ are summarised in Table~\ref{tab:UncSumJetR04_Dzero}. 
The \zch systematic uncertainties in two of the four \ptchjet intervals are presented in Table~\ref{tab:UncSumZR04_Dzero}.
 

\begin{table}[bt]
\caption{Relative (\%) systematic uncertainties for selected \ptchjet intervals of $R=$~0.4 jets at \five and \thirteen.}
\label{tab:UncSumJetR04_Dzero}
\begin{center}
\resizebox{0.75\textwidth}{!}{
\begin{tabular}{l|ccc|ccc}
\multicolumn{1}{r|}{$\sqrt{s}$ (\TeV)}&\multicolumn{3}{c}{$5.02$}&\multicolumn{3}{|c}{$13$}\\
\multicolumn{1}{r|}{\ptchjet (\GeVc)}&$~5$--$6$&$~8$--$10$&$~30$--$50$&$~5$--$6$&$~8$--$10$&$~30$--$50$\\\hline\hline
Topological selection &{\color{white}+~}3.4&{\color{white}+~}5.6&{\color{white}+~}25&{\color{white}+~}3.6&{\color{white}+~}2.9&{\color{white}+~~}8.8\\
Raw yield extraction &{\color{white}+~}3.1&{\color{white}+~}3.1&{\color{white}+~}11&{\color{white}+~}3.3&{\color{white}+~}2.5&{\color{white}+~~}8.8\\
\multirow{2}{*}{B Feed-down}&+~3.9&+~5.3&+~14&+~4.7&+~5.9&+~12\\
&--~6.5&--~8.9&--~24&--~6.5&--~8.5&--~22\\
Unfolding &{\color{white}+~}2.8&{\color{white}+~}0.6&{\color{white}+~}12&{\color{white}+~}2.7&{\color{white}+~}0.7&{\color{white}+~~}0.9\\
Tracking eff. (jet energy scale) &{\color{white}+~}1.6&{\color{white}+~}2.4&{\color{white}+~~}9.6&{\color{white}+~}0.8&{\color{white}+~}2.1&{\color{white}+~~}9.7\\
Tracking eff. (D-meson)&{\color{white}+~}5.0&{\color{white}+~}5.0&{\color{white}+~~}5.0&{\color{white}+~}5.0&{\color{white}+~}5.0&{\color{white}+~~}5.0\\
BR&{\color{white}+~}0.8&{\color{white}+~}0.8&{\color{white}+~~}0.8&{\color{white}+~}0.8&{\color{white}+~}0.8&{\color{white}+~~}0.8\\
Luminosity&{\color{white}+~}2.1&{\color{white}+~}2.1&{\color{white}+~~}2.1&{\color{white}+~}1.7&{\color{white}+~}1.7&{\color{white}+~~}1.7\\\hline
\multirow{ 2}{*}{Total}&+~9.0&+~10&+~35&+~10&+~10&+~21\\
&--~10&--~13&--~40&--~11&--~12&--~28\\
\end{tabular}}
\end{center}
\end{table}
 

\begin{table}[bt]\caption{Relative (\%) systematic uncertainties for selected \zch and \ptchjet intervals of $R=$ 0.4 jets at \five and \thirteen. }
\label{tab:UncSumZR04_Dzero}
\begin{center}
\resizebox{\textwidth}{!}{
\begin{tabular}{l|cccc|cccc}
\multicolumn{1}{r|}{$\sqrt{s}$ (\TeV)}&\multicolumn{4}{c}{$5.02$}&\multicolumn{4}{|c}{$13$}\\
\multicolumn{1}{r|}{\ptchjet (\GeVc)}&\multicolumn{2}{c}{$~5$--$7$}&\multicolumn{2}{c}{$~7$--$10$}&\multicolumn{2}{|c}{$~5$--$7$}&\multicolumn{2}{c}{$~7$--$10$}\\
\multicolumn{1}{r|}{\zch}&$0.6$--$0.7$&$0.9$--$1.0$&$0.6$--$0.7$&$0.9$--$1.0$&$0.6$--$0.7$&$0.9$--$1.0$&$0.6$--$0.7$&$0.9$--$1.0$\\\hline\hline
Topological selection &{\color{white}+~}3.8&{\color{white}+~}2.3&{\color{white}+~}9.2&{\color{white}+~}1.4&{\color{white}+~}2.8&{\color{white}+~}1.7&{\color{white}+~}3.6&{\color{white}+~}1.1\\
Raw yield extraction &{\color{white}+~}3.3&{\color{white}+~}4.3&{\color{white}+~}4.2&{\color{white}+~}2.4&{\color{white}+~}3.5&{\color{white}+~}4.3&{\color{white}+~}5.8&{\color{white}+~}2.4\\
\multirow{2}{*}{B Feed-down}&+~3.0&+~1.9&+~2.5&+~1.2&+~3.4&+~2.8&+~3.4&+~2.2\\
&--~5.1&--~3.2&--~4.2&--~2.8&--~4.8&--~3.8&--~4.9&--~3.3\\
Unfolding &{\color{white}+~}0.4&{\color{white}+~}1.3&{\color{white}+~}0.7&{\color{white}+~}0.0&{\color{white}+~}0.7&{\color{white}+~}0.2&{\color{white}+~}1.6&{\color{white}+~}1.4\\
Tracking eff. (jet energy scale) &{\color{white}+~}1.3&{\color{white}+~}1.7&{\color{white}+~}5.3&{\color{white}+~}3.1&{\color{white}+~}3.2&{\color{white}+~}4.9&{\color{white}+~}3.2&{\color{white}+~}4.9\\\hline
\multirow{ 2}{*}{Total}&+~6.0&+~5.6&+~12&+~4.5&+~6.6&+~7.3&+~8.4&+~6.1\\
&--~7.3&--~6.2&--~12&--~5.0&--~7.4&--~7.8&--~9.1&--~6.6\\
\end{tabular}}
\end{center}
\end{table}


The total systematic uncertainties for the \Dzero-jet \ptchjet-differential cross sections and the \zch distributions were obtained by summing in quadrature the uncertainties estimated for each of the sources.
In the case of cross section ratios for different jet resolution parameters,
the systematic uncertainties due to tracking efficiency of the \Dzero-meson decay products and the normalisation uncertainties are assumed to be fully correlated and, hence, cancel out in the ratios. Systematic uncertainties due to the \Dzero-meson topological selection are partially correlated and an average of the uncertainties for two resolution parameters $R$ was taken. 
Partial correlation was also assumed for the tracking efficiency related to the jet energy scale. A simultaneous-variation method was used to determine the uncertainty, i.e.~the detector response matrices for two given $R$ values were varied simultaneously and the relative uncertainty on the cross section ratio was determined by the difference of the final ratio results obtained with modified and nominal response matrices. 
Systematic uncertainties on the ratio of cross sections for the two colliding energies were obtained by adding them in quadrature, except for the BR uncertainty which was treated as fully correlated. No other correlation was considered given that the data taking periods and the detector conditions were different.


\section{Results}\label{Sec:results}
\subsection{Transverse-momentum differential cross sections}

The \ptchjet-differential cross section of \Dzero jets is defined as
\begin{equation}
    \frac{\text{d}^2\sigma}{\text{d}\ptchjet \text{d}\etajet} (\ptchjet)= \frac{1}{\lumi_{\text{int}}}\frac{1}{\text{BR}} \frac{N(\ptchjet)}{\Delta \etajet \Delta \ptchjet} ,
\label{eq:xsection}
\end{equation}
where $N(\ptchjet)$ is the measured yield in each \ptchjet interval corrected for the acceptance, reconstruction efficiency, b-hadron feed-down contribution, and unfolded for the detector effects. The $\Delta \ptchjet$ is the bin width and $\Delta \etajet=1.8-2R$ is the jet reconstruction acceptance, where $R$ is the jet resolution parameter. Finally, $\lumi_{\text{int}}$ is the integrated luminosity and BR is the branching ratio of the considered \Dzero-meson decay channel.

\begin{figure}[tb!]
    \begin{center}
    \includegraphics[width = 0.9\textwidth]{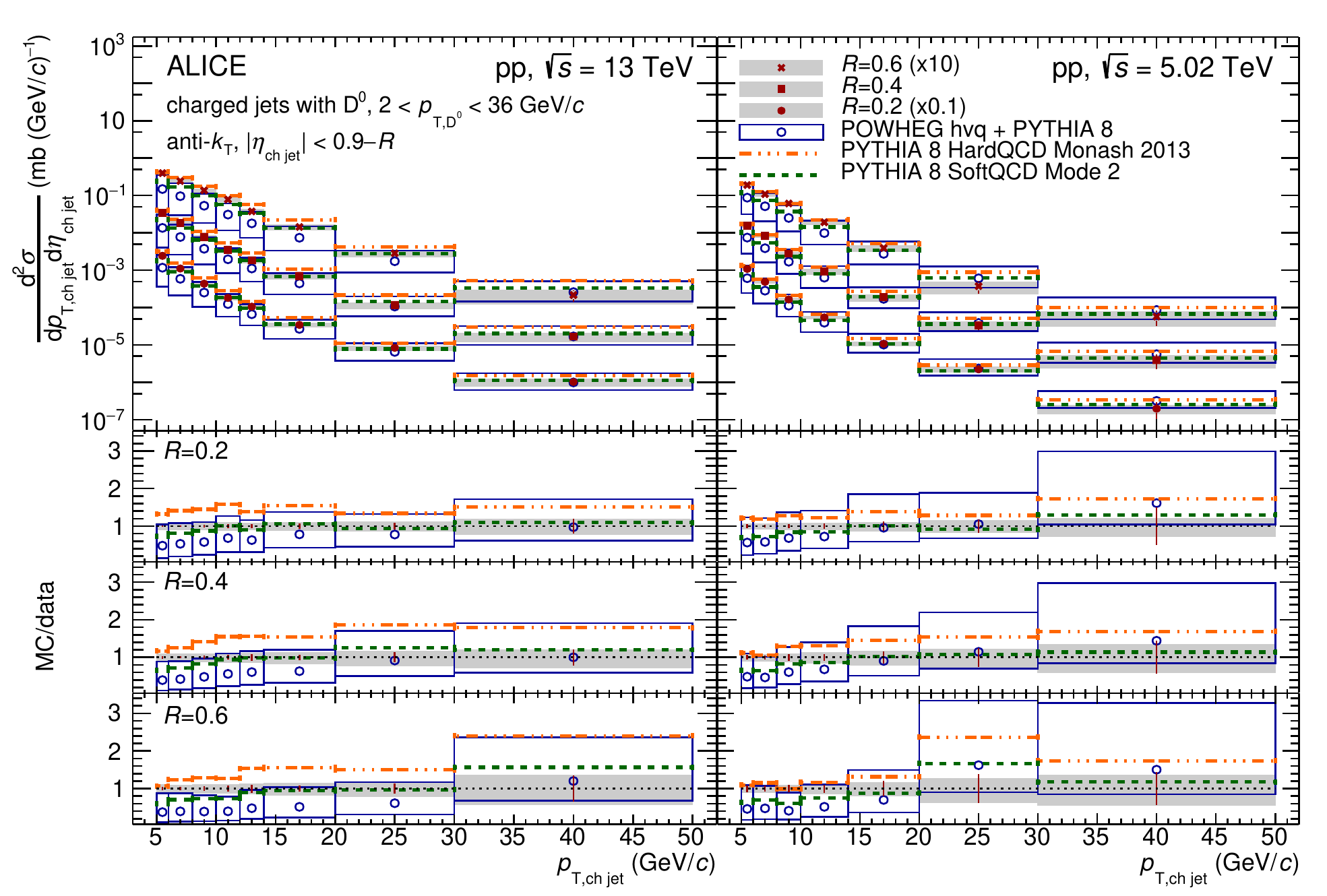}
    \end{center}
    \caption{Top panels: \ptchjet-differential cross section of charm jets tagged with \Dzero mesons for $R=0.2$ (circles, scaled by 0.1), $0.4$ (squares) and $0.6$ (crosses, scaled by 10) in \pp collisions at \thirteen (left) and \five (right) compared to PYTHIA~8 HardQCD Monash~2013 (dash-dotted lines), PYTHIA~8 SoftQCD Mode 2 (dashed lines), and POWHEG hvq + PYTHIA~8 (open circles) predictions. The shaded bands indicate the systematic uncertainty on the data cross section while open boxes represent the theoretical uncertainties on the POWHEG predictions. Bottom panels: ratios of MC predictions to the data for $R=0.2$, $0.4$ and $0.6$.}
    \label{fig:jetptspectra13_5}
\end{figure}

The \ptchjet-differential cross sections of \Dzero jets in \pp
collisions for $R=$ 0.2, 0.4, and 0.6 are shown in Fig.~\ref{fig:jetptspectra13_5} for \thirteen (left) and for \five (right).
They are compared to PYTHIA~8 and POWHEG~+~PYTHIA~8 predictions.
The \ptchjet-differential cross section for $R=$0.3 and its comparisons to theoretical predictions are shown in the appendix in Fig.~\ref{fig:jetspectrar3}.
The jets are required to have in their constituents a \Dzero meson with \ptd~$>$~2~\GeVc as the \Dzero-meson reconstruction efficiency falls rapidly at lower \ptd and excluding  \ptd~$<$~2~\GeVc helps in avoiding large fluctuations in the \ptchjet spectra.
A previous study at \seven~\cite{Acharya:2019zup} showed that a lower bound selection on the \Dzero-meson \pt of \ptd~$>$~3~\GeVc introduced a minimal fragmentation bias on the reported \Dzero-jet \ptchjet-differential cross sections above 5~\GeVc. Therefore, a selection of \ptd~$>$~2~\GeVc should have a smaller effect on the same reported range of \ptchjet spectra.
In this analysis, the maximum transverse momentum was \ptd ~=~ 36~\GeVc for the \Dzero mesons and \ptchjet ~=~ 50~\GeVc for the charged jets.
The same requirements on the \Dzero-meson \pt were applied in the simulations.  

The results are compared to predictions of the Monash-2013 tune~\cite{Sjostrand:2014zea} of the PYTHIA~8.210~\cite{Pythia8} event generator with HardQCD processes. 
It is based on leading order pQCD calculations of matrix elements of parton-level hard scatterings and a leading order parton shower. 
The final state evolution is combined with the initial-state radiation and multiparton interactions. 
The Lund string model~\cite{Andersson:1983ia, Sjostrand:1984ic} is used for the hadronisation. 
It overpredicts the data for all three values of the jet resolution parameter $R$ with the discrepancy being larger at \thirteen.
Incorporating SoftQCD and inelastic non-diffractive processes and colour reconnection beyond the leading-colour approximation~\cite{CR} to the aforementioned PYTHIA~8 tune, denoted as PYTHIA~8 SoftQCD Mode 2, improves the agreement with the data. However, in this case the model underpredicts the measurements at $\ptchjet \lesssim $~10~\GeVc. 

The POWHEG~+~PYTHIA~8 simulation interfaces NLO pQCD POWHEG~\cite{Frixione:2007vw, Nason:2004rx} calculations with the PYTHIA~8~\cite{Pythia8} MC parton shower within the POWHEG Box framework~\cite{Alioli:2010xd}. The heavy-flavour process (hvq)~\cite{Frixione:2007nw} implementation of the POWHEG framework was chosen. The outgoing partons from POWHEG are passed to PYTHIA~8 event-by-event to simulate the subsequent parton shower, hadronisation and generation of the underlying event.
The following simulation settings were used: CT10NLO set of the parton distribution function, the renormalisation and factorisation scales were set to $\mu_{\textrm {R}}$~=~$\mu_{\textrm{F}}$~=~$\mu_0$~=~$\sqrt{m_{\textrm {c}}^2+p_{\textrm {T}}^2}$, and the default charm-quark mass was 1.5 \GeVcsq. 
The theoretical uncertainties were estimated by varying the simulation parameters. 
The largest uncertainties originate from doubling or halving the factorisation and renormalisation scales.
Additionally, the charm-quark mass was varied between 1.3 \GeVcsq and 1.7 \GeVcsq. 
The POWHEG~+~PYTHIA~8 calculations describe the measured cross sections within the experimental and theoretical uncertainties. 
For $\ptchjet >$~14~\GeVc (20~\GeVc) the central values of the predictions agree with the data at \five (\thirteen). 
At lower \ptchjet the experimental results are close to the upper bands of the POWHEG~+~PYTHIA~8 calculations and, as in the case of the PYTHIA~8 predictions, the agreement is better at \five than at \thirteen. 
The low-\ptchjet region is particularly difficult to describe theoretically due to the large contribution from various non-perturbative effects.

\begin{figure}[tb]
    \begin{center}
    \includegraphics[width = 0.95\textwidth]{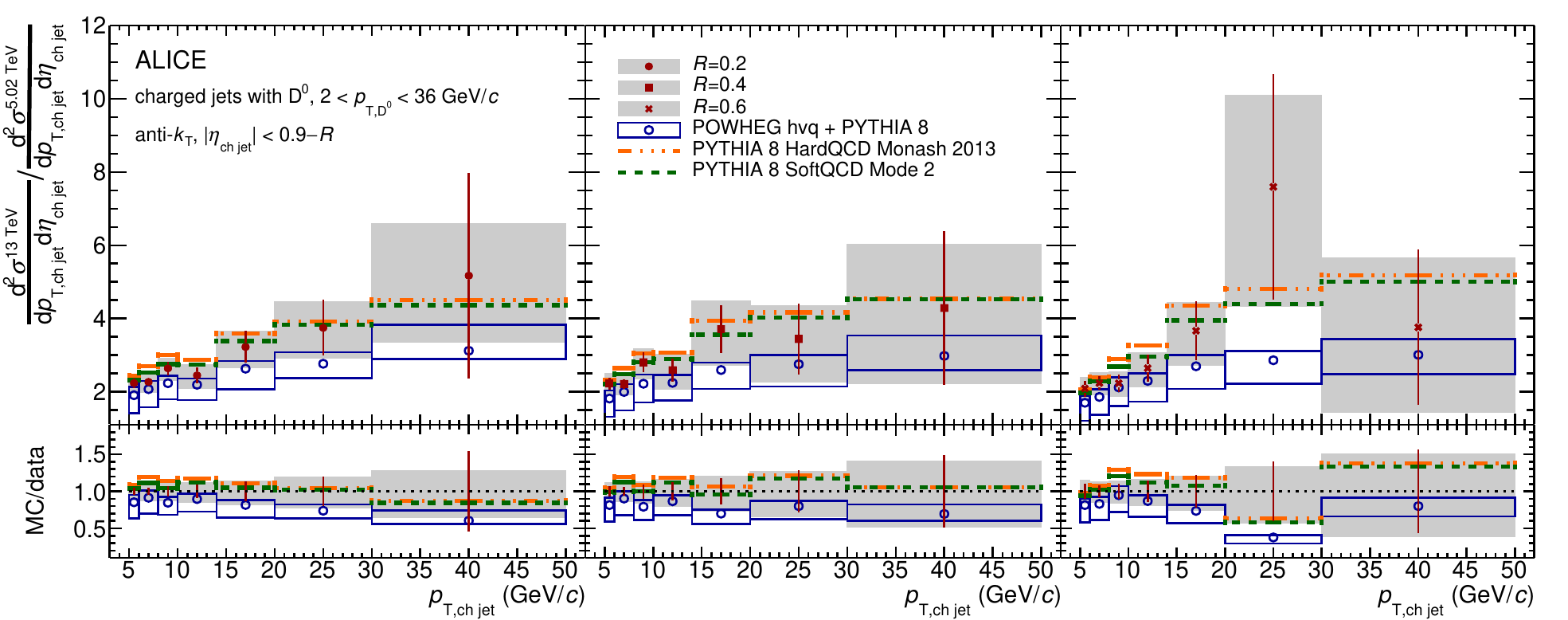}
    \end{center}
    \caption{Top: ratios of \ptchjet-differential cross section of charm jets tagged with \Dzero mesons in \pp collisions at \thirteen to \five for $R=0.2$ (left), $R=0.4$ (centre), and $R=0.6$ (right) compared to PYTHIA~8 HardQCD Monash~2013 (dash-dotted lines), PYTHIA~8 SoftQCD Mode 2 (dashed lines), and POWHEG hvq + PYTHIA~8 (open circles) predictions. The shaded bands indicate the systematic uncertainty on the cross section ratios while open boxes represent the theoretical uncertainties on the POWHEG predictions. 
    Bottom: ratios of MC predictions to the data.
    }
    \label{fig:jetptshowerEnergyRatio}
\end{figure}

In addition, the energy dependence of the \ptchjet-differential cross section of \Dzero jets was studied from the ratio of \thirteen to \five cross sections, shown for different jet resolution parameters $R$ in Fig.~\ref{fig:jetptshowerEnergyRatio}. 
The measured ratios indicate a hardening of the \ptchjet spectra with increasing centre-of-mass energy.
Both PYTHIA~8 settings describe the data well within the current uncertainties for all jet resolution parameters $R$. 
The PYTHIA~8 with SoftQCD and Mode 2 tune describes the data slightly better. 
The POWHEG~+~PYTHIA~8 simulation underestimates the measured cross section ratios, with the data being on the upper edge of the theory uncertainty band.

\subsection{Resolution parameter dependence of \texorpdfstring\Dzero--jet cross section}
\begin{figure}[tb]
    \begin{center}
    \includegraphics[width = 0.9\textwidth]{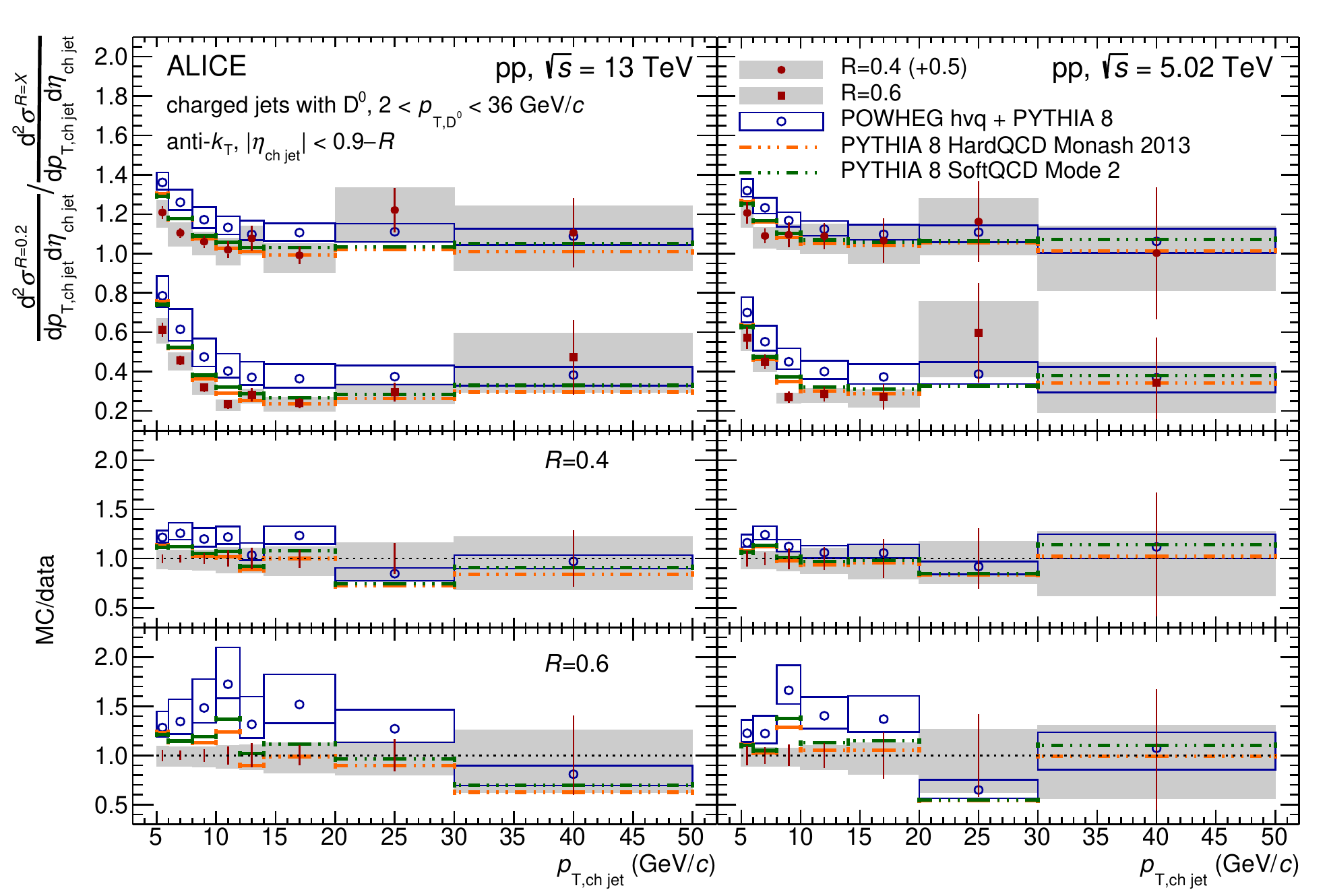}
    \end{center}
    \caption{Top: ratios of \ptchjet-differential cross section of charm jets tagged with \Dzero mesons for different $R$: $\sigma(R=0.2)/\sigma(R=0.4)$ (circles, shifted up by 0.5) and $\sigma(R=0.2)/\sigma(R=0.6)$ (squares) in \pp collisions at \thirteen~(left) and \five~(right) compared to PYTHIA~8 HardQCD Monash~2013 (dash-dotted lines), PYTHIA~8 SoftQCD Mode 2 (dashed lines), and POWHEG hvq + PYTHIA~8 (open circles) predictions. The shaded bands indicate the systematic uncertainty on the cross section ratios while open boxes represent the theoretical uncertainties on the POWHEG predictions.
    Bottom: ratios of MC predictions to the data for \mbox{$\sigma(R=0.2)/\sigma(R=0.4)$} and \mbox{$\sigma(R=0.2)/\sigma(R=0.6)$}, respectively.
    }
    \label{fig:jetptshowerRadiusRatio}
\end{figure}

A comparison of \Dzero jets with different resolution parameters can help in exploring the shower development. 
It provides insights into the interplay between perturbative and non-perturbative effects. 
Figure~\ref{fig:jetptshowerRadiusRatio} shows the ratios of  \ptchjet-differential cross sections of \Dzero jets reconstructed with resolution parameter $R=0.2$ with respect to $R = 0.4$ and $0.6$ for collision energies at \thirteen (left) and \five (right). Statistical uncertainties are treated as fully uncorrelated and summed in quadrature, thus they are overestimated.
To determine the theoretical uncertainties for cross section ratios between two jet radii in the POWHEG framework, the renormalisation and factorisation scales and the charm-quark mass were varied simultaneously. 
The maximum upward and downward variations were used as the uncertainty band.

The observed departure from unity of the cross section ratios can be interpreted by the emission of QCD radiation.
Both $\sigma(R=0.2)/\sigma(R=0.4)$ and $\sigma(R=0.2)/\sigma(R=0.6)$ ratios for the two collision energies decrease with increasing \ptchjet 
and for $\ptchjet > 10$ \GeVc the ratios become independent of \ptchjet within the uncertainties.
The shapes are qualitatively described by the PYTHIA~8 and POWHEG~+~PYTHIA~8 predictions. 

However, in the \ptchjet interval $5$--$10$~\GeVc, POWHEG~+~PYTHIA~8 calculations overestimate the data with the discrepancy being larger for the $\sigma(R=0.2)/\sigma(R=0.6)$ ratio, which is expected to be more sensitive to the underlying event contribution. 
The PYTHIA~8 predictions with the Monash and Mode 2 tunes agree well with the data within the uncertainties, where the largest deviations from the data are at low \ptchjet for \thirteen and $R=$~0.6. 
The differences seen between the predictions of the two PYTHIA~8 tunes in the \ptchjet-differential cross sections largely cancel in the ratios of results with different $R$ parameters.  

\subsection{\texorpdfstring{\Dzero}--jet fraction of inclusive jets}
\begin{figure}[tb!]
    \begin{center}
    \includegraphics[width = \textwidth]{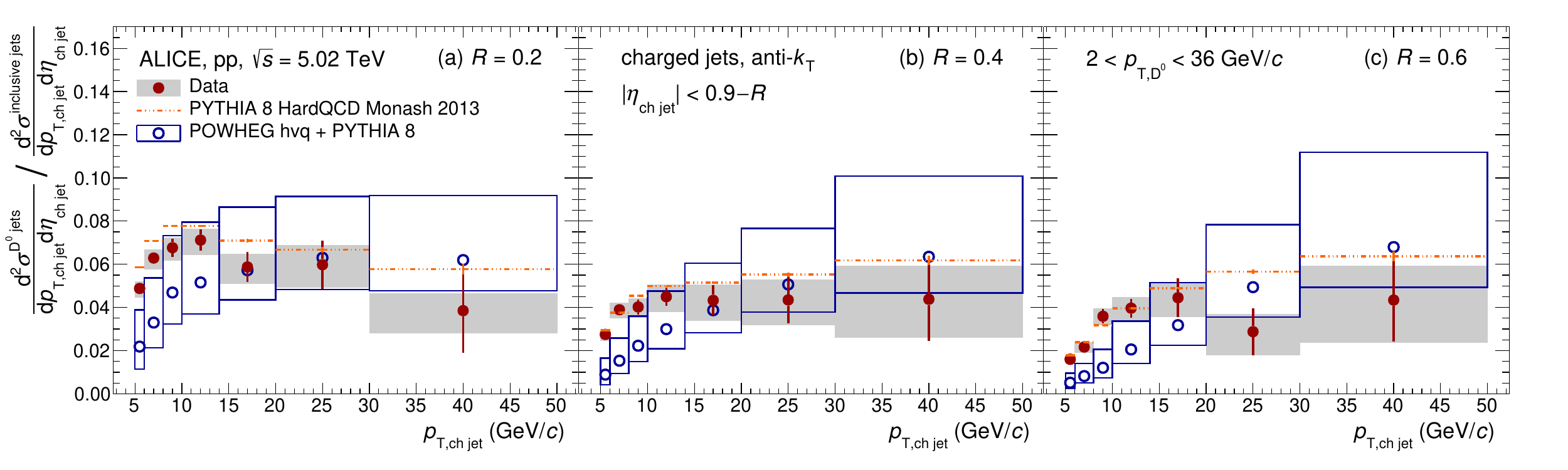}
    \end{center}
    \vspace{-0.2cm}
    \caption{The fraction of \Dzero jets over inclusive charged-particle jets in \pp collisions at \five for \mbox{(a)~$R = 0.2$}, (b)~$R = 0.4$, and (c)~$R = 0.6$ compared to PYTHIA~8 HardQCD Monash~2013 (dash-dotted lines) and POWHEG hvq + PYTHIA~8 (open circles) predictions. 
    The shaded bands indicate the systematic uncertainty on the data cross section ratios while the open boxes represent the theoretical uncertainties on the POWHEG predictions.
    }
    \label{fig:D0jetfraction}
\end{figure}

Figure~\ref{fig:D0jetfraction} shows the fraction of \Dzero jets with respect to charged-particle inclusive jets in \pp collisions at \five for different jet resolution parameters $R= 0.2$, $0.4$, and $0.6$. The production cross sections of the inclusive jets used as a reference here are taken from a previous measurement by ALICE reported in Ref.~\cite{Acharya:2019tku}. Since the data taking periods are different for the inclusive jet measurements compared to the current one, all the uncertainties were considered as uncorrelated.

The fraction of \Dzero jets tends to increase with increasing \ptchjet in the kinematic range 5~$<$~\ptchjet~$<$~10 \GeVc for all jet radii. 
However, the fraction decreases with increasing $R$, from a range of $0.05$--$0.07$ at $R = 0.2$ to a range of $0.015-0.04$ at $R = 0.6$. The \Dzero-jet fraction for $R=0.3$ is shown in the appendix in Fig.~\ref{fig:jetDvInc3}.
In the range of \ptchjet $>$~10~\GeVc, the \ptchjet dependence tends to flatten out within uncertainties due to the hardening of the jets. 
The \Dzero-jet fractions follow the trend set by PYTHIA~8 results with Monash tune and agree with them quite well. The POWHEG~+~PYTHIA~8 calculations slightly underestimate the data at lower \ptchjet while agreeing within uncertainties at higher \ptchjet.
\subsection{Jet momentum fraction}

\begin{figure}[tb]
    \begin{center}
    \includegraphics[width = \textwidth]{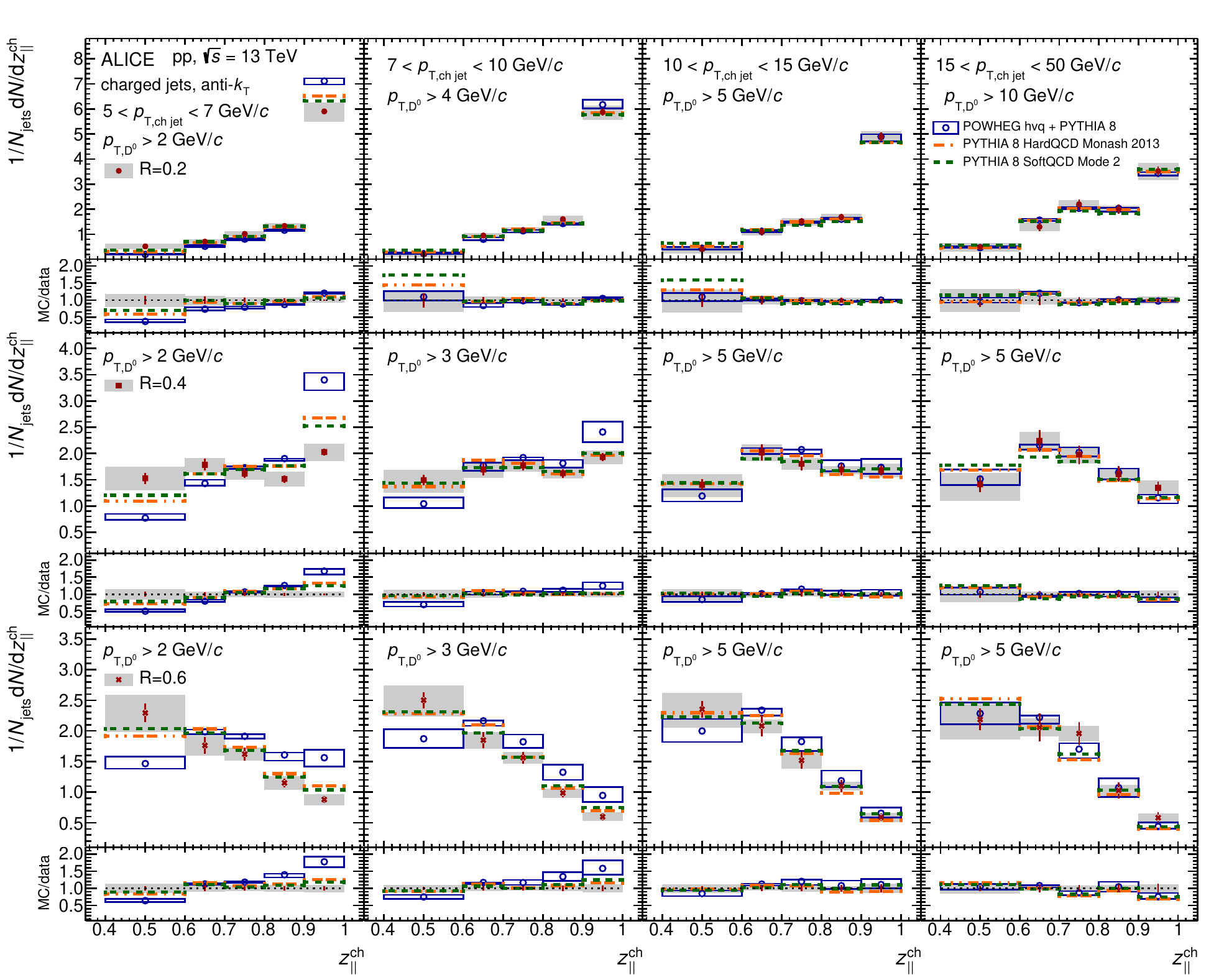}
    \end{center}
    \caption{Distributions of \zch-differential yield of charm jets tagged with \Dzero mesons normalised by the number of \Dzero jets within each distribution in \pp collisions at \thirteen in four jet-\pt intervals ${5<\ptchjet<7~\GeVc}$, ${7<\ptchjet<10~\GeVc}$, ${10<\ptchjet<15~\GeVc}$, and ${15<\ptchjet<50~\GeVc}$ from left to right, respectively. Top, middle, and bottom rows represent jets with $R=0.2$, $0.4$, and $0.6$, respectively. They are compared to PYTHIA~8 HardQCD Monash~2013 (dash-dotted lines), PYTHIA~8 SoftQCD Mode 2 (dashed lines), and POWHEG hvq + PYTHIA~8 (open circles) predictions. The shaded bands indicate the systematic uncertainty on the distributions.
    Bottom panels present ratios of MC predictions to the data.}
    \label{fig:zch13}
\end{figure}

\begin{figure}[tb]
    \begin{center}
    \includegraphics[width = \textwidth]{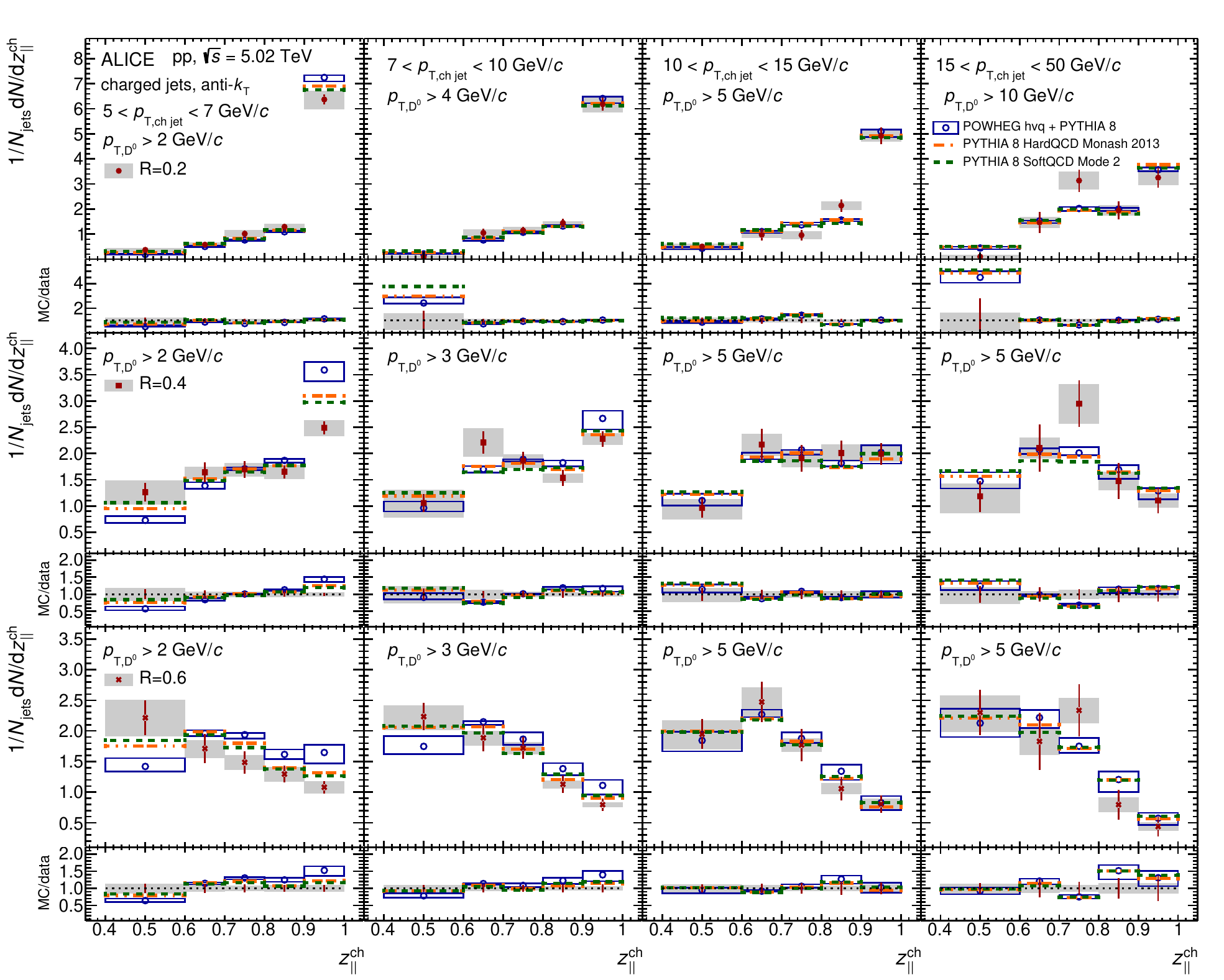}
    \end{center}
    \caption{Distributions of \zch-differential yield of charm jets tagged with \Dzero mesons and normalised by the number of \Dzero jets within each distribution in \pp collisions at \five in four \ptchjet intervals ${5<\ptchjet<7~\GeVc}$, ${7<\ptchjet<10~\GeVc}$, ${10<\ptchjet<15~\GeVc}$, and ${15<\ptchjet<50~\GeVc}$ from left to right, respectively. Top, middle, and bottom rows represent jets with $R=0.2$, $0.4$, and $0.6$, respectively. They are compared to PYTHIA~8 HardQCD Monash~2013 (dash-dotted lines), PYTHIA~8 SoftQCD Mode 2 (dashed lines), and POWHEG hvq + PYTHIA~8 (open circles) predictions. The shaded bands indicate the systematic uncertainty on the distributions.
    Bottom panels present ratios of MC predictions to the data.}
    \label{fig:zch5}
\end{figure}

The fraction of jet momentum carried by the \Dzero meson can provide insight into the charm-quark fragmentation. 
The \zch-differential yield, $\text{d}^2N/\text{d}\zch \text{d}\etajet$, was calculated in a manner analogous to the calculation of \ptchjet-differential cross section (see~\ref{eq:xsection}). 
It was then self-normalised in each \ptchjet interval 
by the integral of the measured \zch distribution in the corresponding \ptchjet interval to obtain the presented \zch probability density distributions
\begin{equation}
    \frac{1}{N} \frac{\text{d}^2 N}{\text{d}\zch \text{d}\etajet} (\zch, \ptchjet) = \frac{1}{N (\ptchjet)} \frac{N(\zch, \ptchjet)}{\Delta \etajet \Delta \zch}.
\label{eq:zchProb}
\end{equation}
This normalisation was applied in order to better compare the shape of the distributions among each other and to different model predictions. 
Figures~\ref{fig:zch13} and~\ref{fig:zch5} show the \zch distributions in four different intervals of \ptchjet for \thirteen and \five, respectively. 
The distributions for $R=0.3$ \mbox{\Dzero jets} at \five are shown in Fig.~\ref{fig:zch53}.
A \ptchjet-dependent minimum \Dzero-meson \pt requirement had to be applied due to the limited number of candidates in some momentum intervals.
For $R=0.2$, these were \ptd~$>$~2, 4, 5, and 10 \GeVc in the \ptchjet ranges $5<\ptchjet<7$, $7<\ptchjet<10$, $10<\ptchjet<15$, and $15<\ptchjet<50$~\GeVc, respectively. For $R=$~0.4 and 0.6, the respective required selections on the minimum \ptd were: 2, 3, 5, and 5~\GeVc. The same kinematic conditions were adopted in the model calculations. 

For \Dzero jets with 5~$<$~\ptchjet $<$~15~\GeVc and reconstructed with $R=$~0.2, a peak at $\zch \approx$~1 is visible, for both \five and \thirteen. 
The peak contains jets whose only constituent is the tagged \Dzero meson and it disappears at larger $R$ and higher \ptchjet intervals where the fraction of these single-constituent jets becomes much smaller.
For a given \ptchjet interval, a softening of the fragmentation (\zch) is visible with increasing $R$. 
The change in the \zch distribution shape with increasing \ptchjet is significant only for $R=$~0.4, with a trend that is similar to that reported in previous ALICE studies at \seven~\cite{Acharya:2019zup}.

The measured \zch distributions are compared to the predictions of the same models used for the \ptchjet-differential cross section.
Overall, a good agreement between PYTHIA~8 results with both Monash and Mode 2 tunes and the data is observed within the uncertainties for 7~$<$ \ptchjet $<$~50~\GeVc at both collision energies. A hint of a softer fragmentation in the lowest \ptchjet interval, 5~$<$~\ptchjet $<$~7~\GeVc, is visible in the data compared to the PYTHIA~8 predictions. The differences in the \zch distribution shape predicted by the default PYTHIA~8 Monash 2013 tune and the SoftQCD Mode 2 are very small, with a slightly harder fragmentation predicted by the former at low \ptchjet and smaller $R$.
Similarly, POWHEG~+~PYTHIA~8 describes the data well above $\ptchjet > 7$ (10) \GeVc at \five (\thirteen) while it predicts a harder fragmentation at lower \ptchjet. The discrepancy between the data and POWHEG~+~PYTHIA~8 predictions in these lower \ptchjet ranges in the \zch distribution shape is larger than in the case of the PYTHIA~8 event generator. It is particularly significant at \thirteen in the interval 5~$<$~\ptchjet $<$~10~\GeVc for jets reconstructed with $R=$~0.6 and 0.4 and for 5~$< \ptchjet <$~7~\GeVc with $R=$~0.2. The discrepancy is larger for larger $R$.

\section{Summary}\label{sec:summary}
In this paper, studies of the production of charm jets tagged with fully reconstructed \Dzero mesons, using data obtained from proton--proton collisions at \five and \thirteen with the ALICE detector at the CERN LHC, were presented. 
The measurements were carried out for charged-particle jets reconstructed with different resolution parameters, i.e.~$R = 0.2, 0.4, 0.6$. 
The new ALICE results shown in this paper have better precision for the studied observables and are performed  more differentially owing to larger data samples of pp collisions at $\s~=~13$ and 5.02~TeV collected by ALICE compared to the results obtained at \seven~\cite{Acharya:2019zup}. 
They are differential in \ptchjet and double differential in \zch and \ptchjet, 
and are compared to predictions obtained with the PYTHIA~8 event generator with the Monash tune as well as with the Mode 2 tune (implementing colour reconnection beyond the leading-colour approximation), and to predictions obtained by coupling the POWHEG NLO event generator to the PYTHIA~8 parton shower. 

The PYTHIA~8 predictions with the SoftQCD and Mode 2 tune settings provide the best description of the \ptchjet-differential cross sections for both collision energies and all resolution parameters. 
Within the experimental and theoretical uncertainties, the measurements are also in agreement with the POWHEG~+~PYTHIA~8 calculations.
Cross section ratios between $\s~=~13$ and 5.02~TeV increase with increasing \ptchjet, indicating a hardening of the spectrum as the collision energy rises. 
The cross section ratios between different jet radii $\sigma(R=0.2)/\sigma(R=0.4,0.6)$ fall sharply with \ptchjet and then flatten out for $\ptchjet >$~10~\GeVc.
Low-\ptchjet measurements for different $R$ values can constrain pQCD, hadronisation, and underlying event (UE) effects in models. 
Studies for smaller $R$ values are more sensitive to non-perturbative hadronisation effects, while contributions from the UE are more important for large $R$.
The ratios are well described by the PYTHIA~8 predictions and are systematically overpredicted by the POWHEG~+~PYTHIA~8 calculations, especially for $\ptchjet < $~20~\GeVc and \thirteen.

The probability density distributions of the jet momentum fraction carried by the constituent \Dzero meson, \zch, hint at a softer fragmentation in data when compared to model predictions in the low \ptchjet region and for larger jet radii. 
This disagreement is more prominent for NLO predictions obtained from POWHEG~+~PYTHIA~8 than PYTHIA~8 predictions. For $\ptchjet > $~7~\GeVc, the agreement between data and the calculations is good.

Despite these discrepancies at low \ptchjet, a generally good description of the main features of the data is obtained with MC event generators and pQCD calculations in most of the measured kinematic range, indicating that the charm-quark production, fragmentation and hadronisation are under control. Therefore, these models can serve as a good theoretical baseline for studies in p--Pb and Pb--Pb collisions.
The reported \zch distributions also serve as an important input for the global fit analyses that aim to constrain the gluon fragmentation functions.  
Furthermore, the results from \pp collisions at \five are at the same centre-of-mass energy as p--Pb and Pb--Pb collision data and can be used as a reference for studies of charm-jet production and fragmentation modifications in the QGP medium and cold nuclear matter effects in p--Pb collisions.


\newenvironment{acknowledgement}{\relax}{\relax}
\begin{acknowledgement}
\section*{Acknowledgements}

The ALICE Collaboration would like to thank all its engineers and technicians for their invaluable contributions to the construction of the experiment and the CERN accelerator teams for the outstanding performance of the LHC complex.
The ALICE Collaboration gratefully acknowledges the resources and support provided by all Grid centres and the Worldwide LHC Computing Grid (WLCG) collaboration.
The ALICE Collaboration acknowledges the following funding agencies for their support in building and running the ALICE detector:
A. I. Alikhanyan National Science Laboratory (Yerevan Physics Institute) Foundation (ANSL), State Committee of Science and World Federation of Scientists (WFS), Armenia;
Austrian Academy of Sciences, Austrian Science Fund (FWF): [M 2467-N36] and Nationalstiftung f\"{u}r Forschung, Technologie und Entwicklung, Austria;
Ministry of Communications and High Technologies, National Nuclear Research Center, Azerbaijan;
Conselho Nacional de Desenvolvimento Cient\'{\i}fico e Tecnol\'{o}gico (CNPq), Financiadora de Estudos e Projetos (Finep), Funda\c{c}\~{a}o de Amparo \`{a} Pesquisa do Estado de S\~{a}o Paulo (FAPESP) and Universidade Federal do Rio Grande do Sul (UFRGS), Brazil;
Bulgarian Ministry of Education and Science, within the National Roadmap for Research Infrastructures 2020-2027 (object CERN), Bulgaria;
Ministry of Education of China (MOEC) , Ministry of Science \& Technology of China (MSTC) and National Natural Science Foundation of China (NSFC), China;
Ministry of Science and Education and Croatian Science Foundation, Croatia;
Centro de Aplicaciones Tecnol\'{o}gicas y Desarrollo Nuclear (CEADEN), Cubaenerg\'{\i}a, Cuba;
Ministry of Education, Youth and Sports of the Czech Republic, Czech Republic;
The Danish Council for Independent Research | Natural Sciences, the VILLUM FONDEN and Danish National Research Foundation (DNRF), Denmark;
Helsinki Institute of Physics (HIP), Finland;
Commissariat \`{a} l'Energie Atomique (CEA) and Institut National de Physique Nucl\'{e}aire et de Physique des Particules (IN2P3) and Centre National de la Recherche Scientifique (CNRS), France;
Bundesministerium f\"{u}r Bildung und Forschung (BMBF) and GSI Helmholtzzentrum f\"{u}r Schwerionenforschung GmbH, Germany;
General Secretariat for Research and Technology, Ministry of Education, Research and Religions, Greece;
National Research, Development and Innovation Office, Hungary;
Department of Atomic Energy Government of India (DAE), Department of Science and Technology, Government of India (DST), University Grants Commission, Government of India (UGC) and Council of Scientific and Industrial Research (CSIR), India;
National Research and Innovation Agency - BRIN, Indonesia;
Istituto Nazionale di Fisica Nucleare (INFN), Italy;
Japanese Ministry of Education, Culture, Sports, Science and Technology (MEXT) and Japan Society for the Promotion of Science (JSPS) KAKENHI, Japan;
Consejo Nacional de Ciencia (CONACYT) y Tecnolog\'{i}a, through Fondo de Cooperaci\'{o}n Internacional en Ciencia y Tecnolog\'{i}a (FONCICYT) and Direcci\'{o}n General de Asuntos del Personal Academico (DGAPA), Mexico;
Nederlandse Organisatie voor Wetenschappelijk Onderzoek (NWO), Netherlands;
The Research Council of Norway, Norway;
Commission on Science and Technology for Sustainable Development in the South (COMSATS), Pakistan;
Pontificia Universidad Cat\'{o}lica del Per\'{u}, Peru;
Ministry of Education and Science, National Science Centre and WUT ID-UB, Poland;
Korea Institute of Science and Technology Information and National Research Foundation of Korea (NRF), Republic of Korea;
Ministry of Education and Scientific Research, Institute of Atomic Physics, Ministry of Research and Innovation and Institute of Atomic Physics and University Politehnica of Bucharest, Romania;
Ministry of Education, Science, Research and Sport of the Slovak Republic, Slovakia;
National Research Foundation of South Africa, South Africa;
Swedish Research Council (VR) and Knut \& Alice Wallenberg Foundation (KAW), Sweden;
European Organization for Nuclear Research, Switzerland;
Suranaree University of Technology (SUT), National Science and Technology Development Agency (NSTDA), Thailand Science Research and Innovation (TSRI) and National Science, Research and Innovation Fund (NSRF), Thailand;
Turkish Energy, Nuclear and Mineral Research Agency (TENMAK), Turkey;
National Academy of  Sciences of Ukraine, Ukraine;
Science and Technology Facilities Council (STFC), United Kingdom;
National Science Foundation of the United States of America (NSF) and United States Department of Energy, Office of Nuclear Physics (DOE NP), United States of America.
In addition, individual groups or members have received support from:
Marie Sk\l{}odowska Curie, Strong 2020 - Horizon 2020, European Research Council (grant nos. 824093, 896850, 950692), European Union;
Academy of Finland (Center of Excellence in Quark Matter) (grant nos. 346327, 346328), Finland;
Programa de Apoyos para la Superaci\'{o}n del Personal Acad\'{e}mico, UNAM, Mexico.

\end{acknowledgement}

\bibliographystyle{utphys}   
\bibliography{bibliography}

\newpage
\appendix

\section{Measurements of \texorpdfstring{$\mathrm{D^0}$}{Dzero} jets with \emph{R} = 0.3 in pp collisions at \texorpdfstring{$\sqrt{s}~=~5.02$~TeV}{sqrt(s)=5 TeV}}
\begin{figure}[b!]
    \centering
        \centering
        \includegraphics[width = 0.55\textwidth]{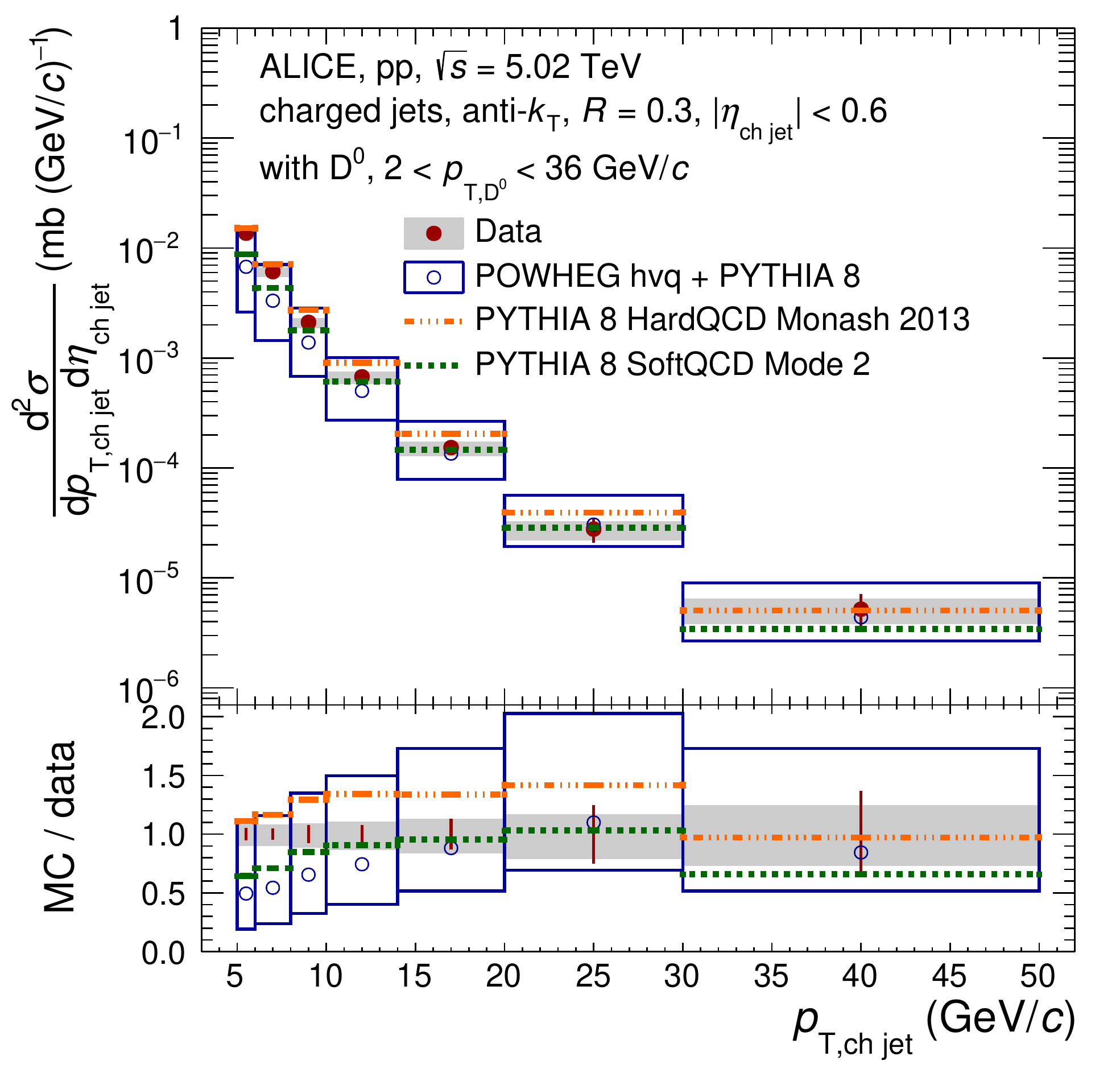}
    \caption{Top panel: \ptchjet-differential cross section of charm jets tagged with \Dzero mesons for $R=0.3$ in \pp collisions at \five compared to PYTHIA~8 HardQCD Monash~2013 (dash-dotted lines), PYTHIA~8 Monash~2013 SoftQCD Mode 2 (dashed lines) and POWHEG hvq + PYTHIA~8 (open circles) predictions. The shaded bands indicate the systematic uncertainty on the data cross section while open boxes represent the theoretical uncertainties on the POWHEG predictions. Bottom panel presents ratios of MC predictions to the data.}
    \label{fig:jetspectrar3}
\end{figure}

\begin{figure}[b!]
    \centering
        \includegraphics[width = 0.55\textwidth]{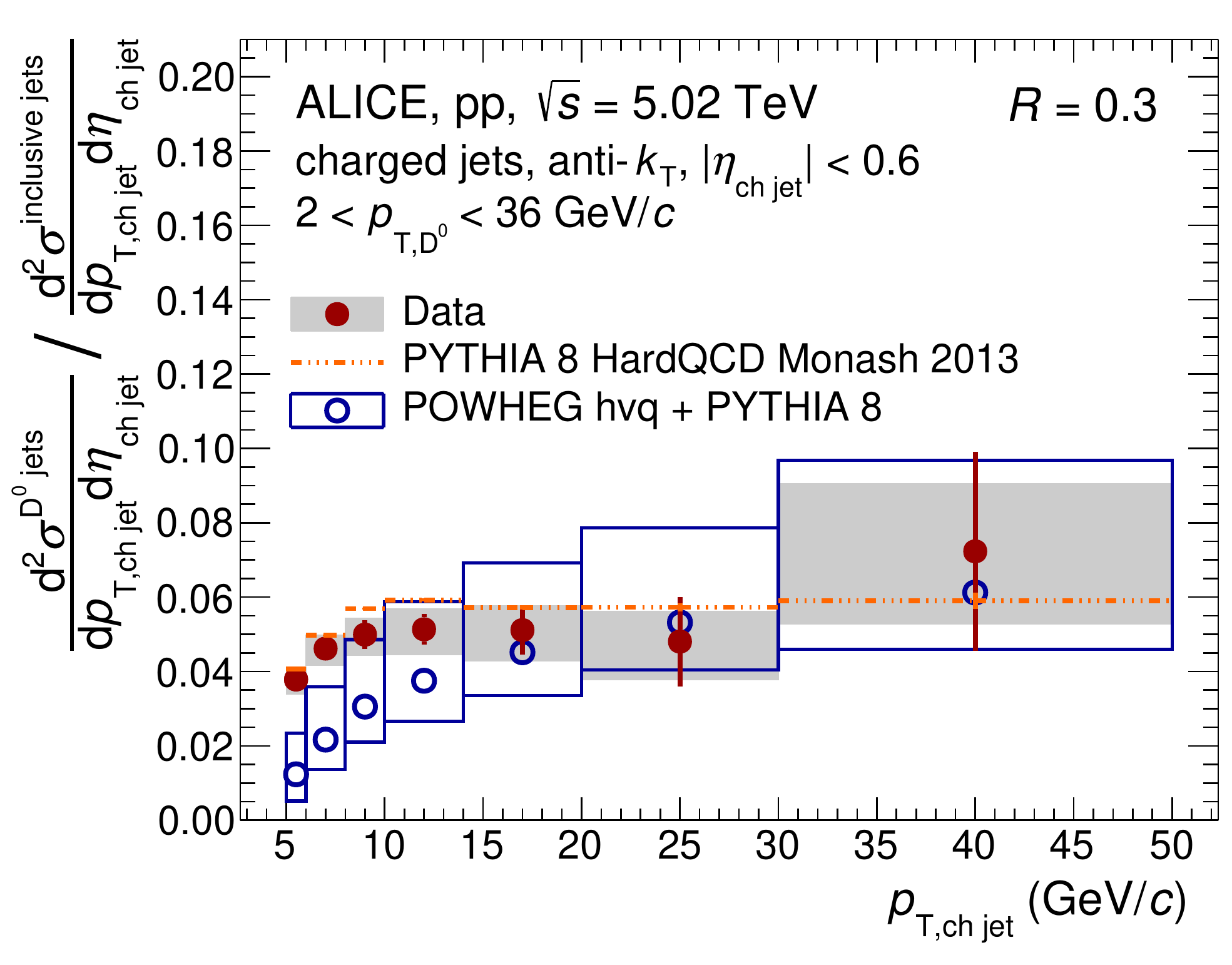}
    \caption{The fraction of \Dzero jets over inclusive charged-particle jets in \pp collisions at \five for \mbox{$R = 0.3$} compared to PYTHIA~8 HardQCD Monash~2013 (dash-dotted lines) and POWHEG hvq + PYTHIA~8 (open circles) predictions.
    }
    \label{fig:jetDvInc3}
\end{figure}

\begin{figure}[tb!]
    \centering
    \begin{minipage}[tbh]{0.45\textwidth}
        \centering
        \includegraphics[width=\textwidth]{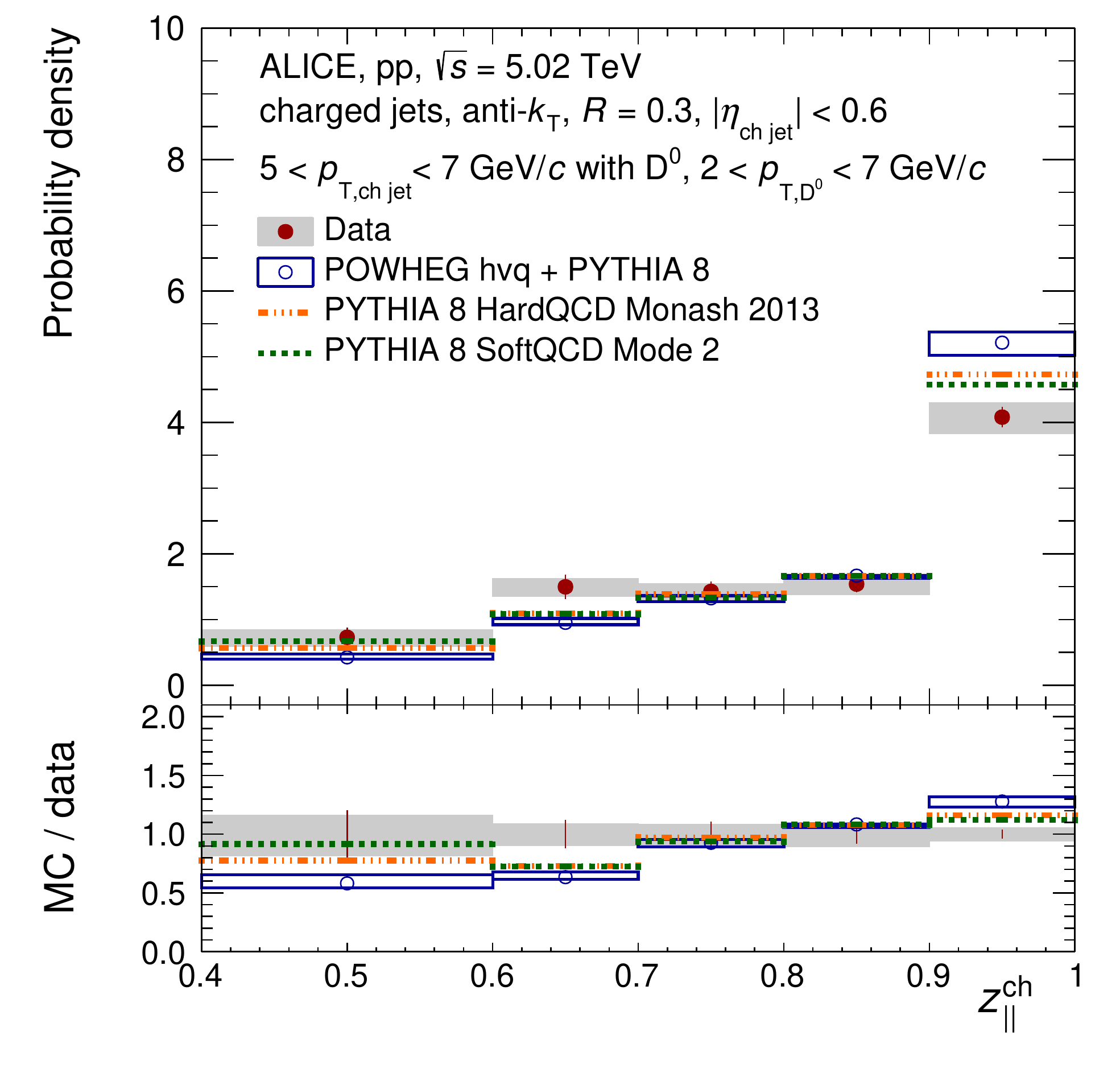}
    \end{minipage}
    \begin{minipage}[tbh]{0.45\textwidth}
        \centering
        \includegraphics[width=\textwidth]{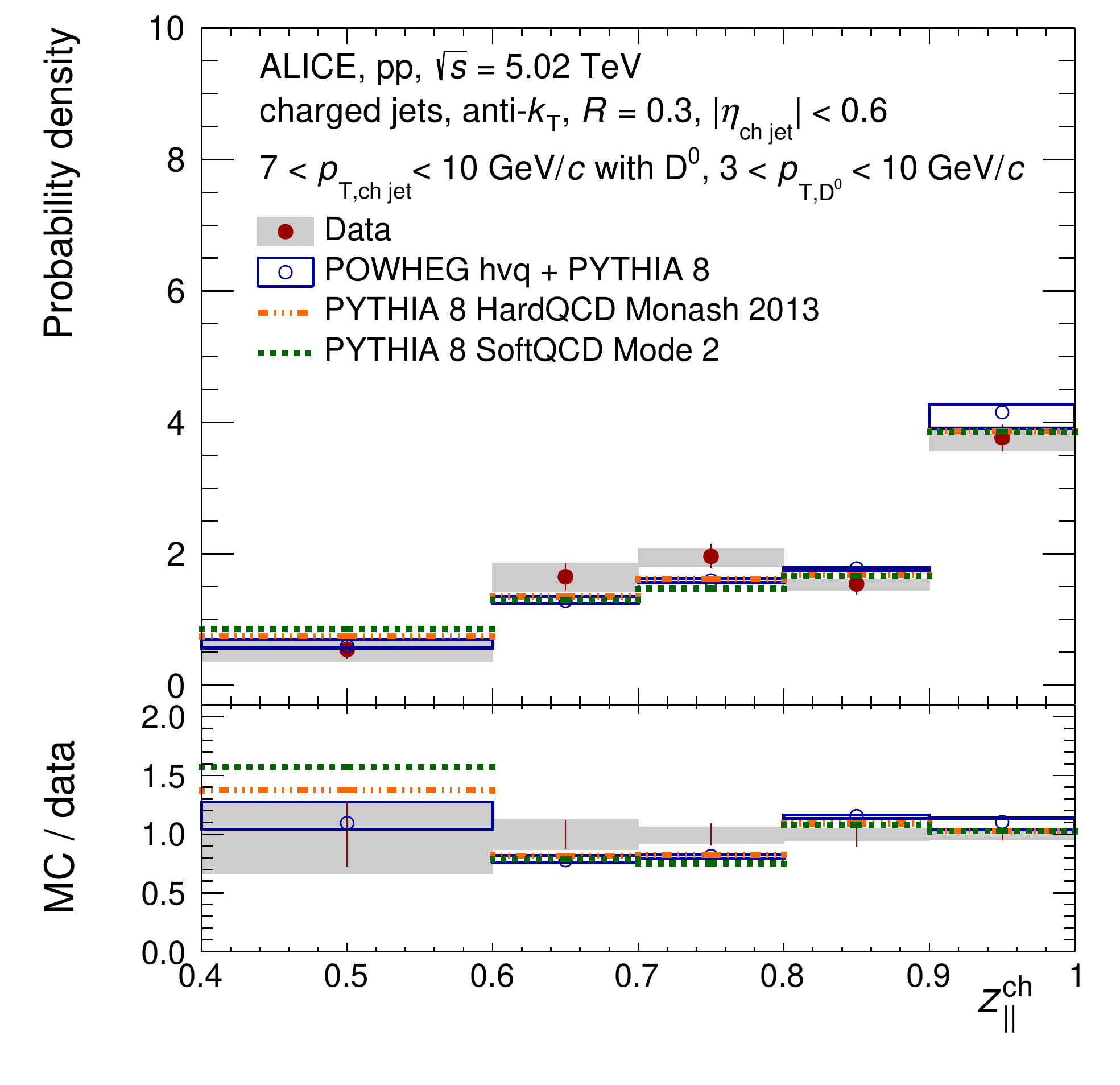}
    \end{minipage}
    \begin{minipage}[tbh]{0.45\textwidth}
        \centering
        \includegraphics[width=\textwidth]{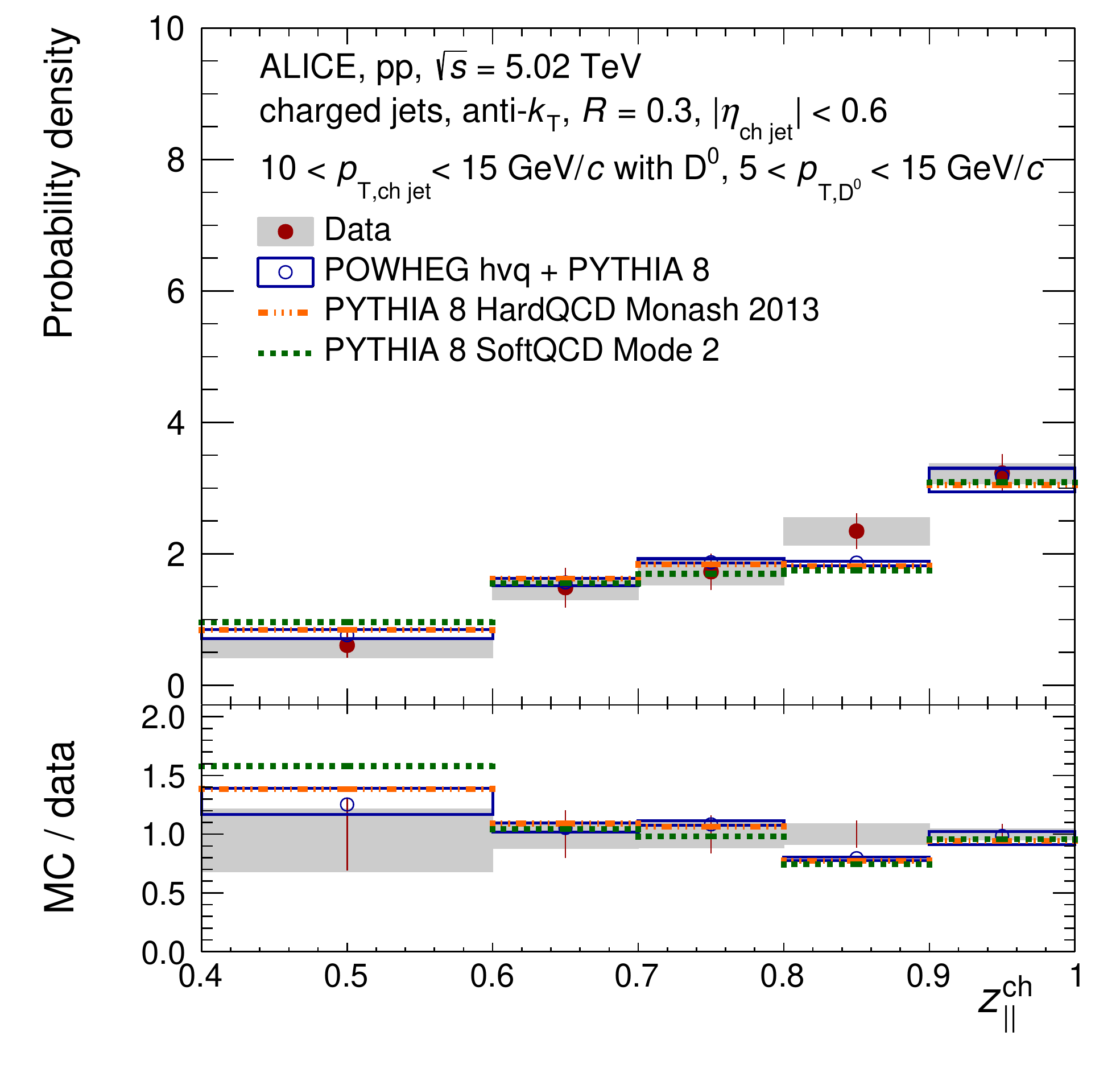}
    \end{minipage}
    \begin{minipage}[tbh]{0.45\textwidth}
        \centering
        \includegraphics[width=\textwidth]{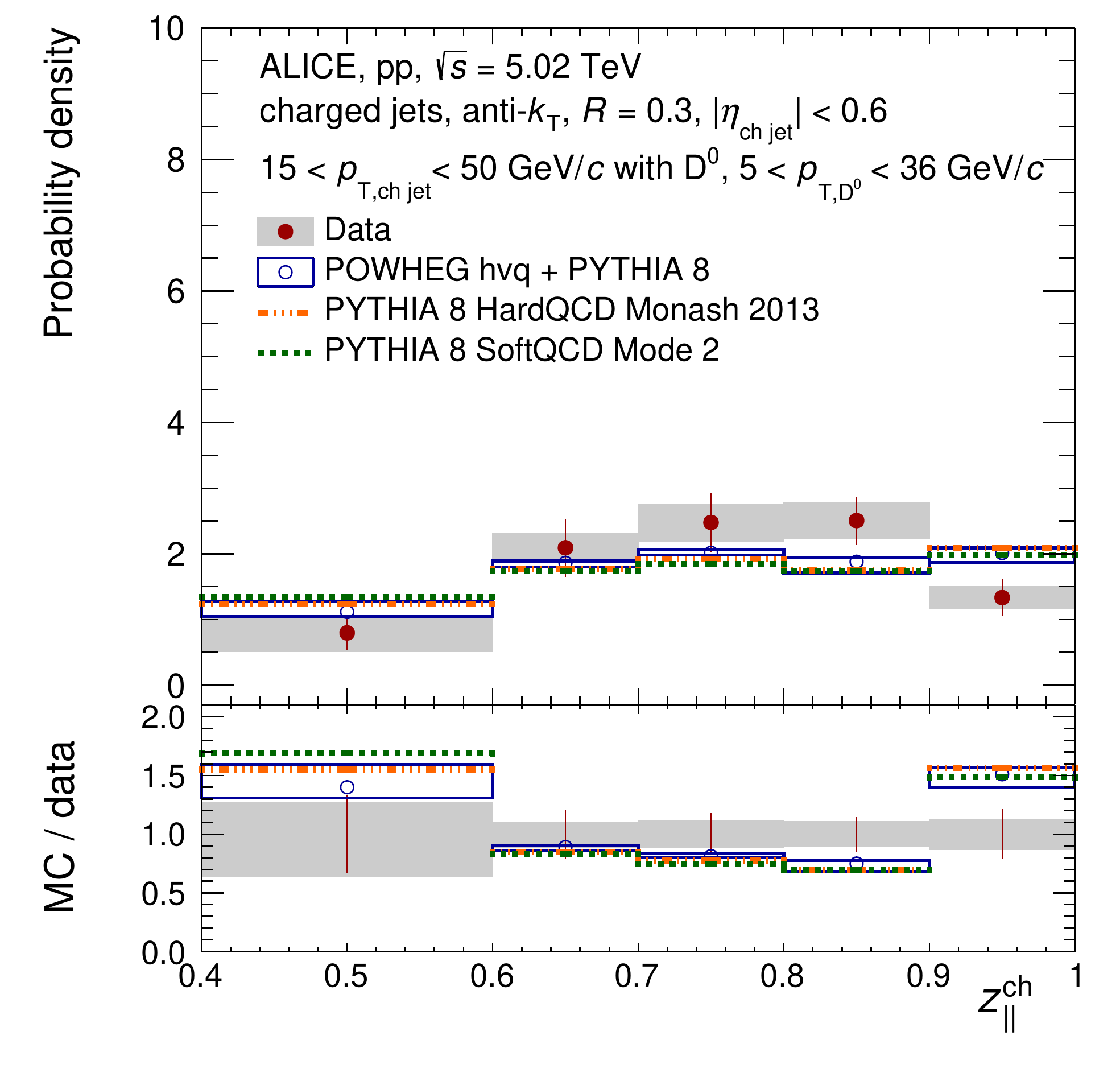}
    \end{minipage}
    \caption{Top panels: \zch-differential yield of $R=0.3$  charm jets tagged with \Dzero\ mesons normalised by the number of \Dzero jets within each distribution in \pp collisions at \five in four \ptchjet intervals (top left) ${5<\ptchjet<7~\GeVc}$, (top right) ${7<\ptchjet<10~\GeVc}$, (bottom left) ${10<\ptchjet<15~\GeVc}$ and (bottom right) ${15<\ptchjet<50~\GeVc}$ \GeVc. They are compared to PYTHIA~8 Monash~2013 (dashed-dotted lines), PYTHIA~8 Monash~2013 SoftQCD Mode 2 (dashed lines) and POWHEG hvq + PYTHIA~8 (open circles) predictions. The shaded bands indicate the systematic uncertainty on the distributions while open boxes represent the theoretical uncertainties on the POWHEG predictions. Bottom panels present ratios of MC predictions to the data.}
    \label{fig:zch53}
\end{figure}

The \ptchjet-differential cross section of \Dzero jets with $R=0.3$ in \pp
collisions at \five compared to PYTHIA~8 and POWHEG+PYTHIA~8 predictions is shown in Fig.~\ref{fig:jetspectrar3}. The \Dzero-jet fraction of inclusive jets for the same $R$ is shown in Fig.~\ref{fig:jetDvInc3}.
Fig.~\ref{fig:zch53} shows the \zch\ distributions for $R=0.3$ \mbox{\Dzero jets} in four different intervals of \ptchjet\ for \five.

\clearpage
%
%

\section{The ALICE Collaboration}
\label{app:collab}
\begin{flushleft} 
\small

S.~Acharya\,\orcidlink{0000-0002-9213-5329}\,$^{\rm 124,131}$, 
D.~Adamov\'{a}\,\orcidlink{0000-0002-0504-7428}\,$^{\rm 86}$, 
A.~Adler$^{\rm 69}$, 
G.~Aglieri Rinella\,\orcidlink{0000-0002-9611-3696}\,$^{\rm 32}$, 
M.~Agnello\,\orcidlink{0000-0002-0760-5075}\,$^{\rm 29}$, 
N.~Agrawal\,\orcidlink{0000-0003-0348-9836}\,$^{\rm 50}$, 
Z.~Ahammed\,\orcidlink{0000-0001-5241-7412}\,$^{\rm 131}$, 
S.~Ahmad\,\orcidlink{0000-0003-0497-5705}\,$^{\rm 15}$, 
S.U.~Ahn\,\orcidlink{0000-0001-8847-489X}\,$^{\rm 70}$, 
I.~Ahuja\,\orcidlink{0000-0002-4417-1392}\,$^{\rm 37}$
% ...

\section*{Affiliation Notes}

$^{\rm I}$ Deceased\\
$^{\rm II}$ Also at: Max-Planck-Institut f\"{u}r Physik, Munich, Germany\\
$^{\rm III}$ Also at: Italian National Agency for New Technologies, Energy and Sustainable Economic Development (ENEA), Bologna, Italy\\
$^{\rm IV}$ Also at: Dipartimento DET del Politecnico di Torino, Turin, Italy\\
$^{\rm V}$ Also at: Department of Applied Physics, Aligarh Muslim University, Aligarh, India\\
$^{\rm VI}$ Also at: Institute of Theoretical Physics, University of Wroclaw, Poland\\
$^{\rm VII}$ Also at: An institution covered by a cooperation agreement with CERN\\

\section*{Collaboration Institutes}

$^{1}$ A.I. Alikhanyan National Science Laboratory (Yerevan Physics Institute) Foundation, Yerevan, Armenia\\
$^{2}$ AGH University of Science and Technology, Cracow, Poland\\
$^{3}$ Bogolyubov Institute for Theoretical Physics, National Academy of Sciences of Ukraine, Kiev, Ukraine\\
$^{4}$ Bose Institute, Department of Physics  and Centre for Astroparticle Physics and Space Science (CAPSS), Kolkata, India\\
$^{5}$ California Polytechnic State University, San Luis Obispo, California, United States\\
$^{6}$ Central China Normal University, Wuhan, China\\
$^{7}$ Centro de Aplicaciones Tecnol\'{o}gicas y Desarrollo Nuclear (CEADEN), Havana, Cuba\\
$^{8}$ Centro de Investigaci\'{o}n y de Estudios Avanzados (CINVESTAV), Mexico City and M\'{e}rida, Mexico\\
$^{9}$ Chicago State University, Chicago, Illinois, United States\\
$^{10}$ China Institute of Atomic Energy, Beijing, China\\
$^{11}$ Chungbuk National University, Cheongju, Republic of Korea\\
$^{12}$ Comenius University Bratislava, Faculty of Mathematics, Physics and Informatics, Bratislava, Slovak Republic\\
$^{13}$ COMSATS University Islamabad, Islamabad, Pakistan\\
$^{14}$ Creighton University, Omaha, Nebraska, United States\\
$^{15}$ Department of Physics, Aligarh Muslim University, Aligarh, India\\
$^{16}$ Department of Physics, Pusan National University, Pusan, Republic of Korea\\
$^{17}$ Department of Physics, Sejong University, Seoul, Republic of Korea\\
$^{18}$ Department of Physics, University of California, Berkeley, California, United States\\
$^{19}$ Department of Physics, University of Oslo, Oslo, Norway\\
$^{20}$ Department of Physics and Technology, University of Bergen, Bergen, Norway\\
$^{21}$ Dipartimento di Fisica, Universit\`{a} di Pavia, Pavia, Italy\\
$^{22}$ Dipartimento di Fisica dell'Universit\`{a} and Sezione INFN, Cagliari, Italy\\
$^{23}$ Dipartimento di Fisica dell'Universit\`{a} and Sezione INFN, Trieste, Italy\\
$^{24}$ Dipartimento di Fisica dell'Universit\`{a} and Sezione INFN, Turin, Italy\\
$^{25}$ Dipartimento di Fisica e Astronomia dell'Universit\`{a} and Sezione INFN, Bologna, Italy\\
$^{26}$ Dipartimento di Fisica e Astronomia dell'Universit\`{a} and Sezione INFN, Catania, Italy\\
$^{27}$ Dipartimento di Fisica e Astronomia dell'Universit\`{a} and Sezione INFN, Padova, Italy\\
$^{28}$ Dipartimento di Fisica `E.R.~Caianiello' dell'Universit\`{a} and Gruppo Collegato INFN, Salerno, Italy\\
$^{29}$ Dipartimento DISAT del Politecnico and Sezione INFN, Turin, Italy\\
$^{30}$ Dipartimento di Scienze MIFT, Universit\`{a} di Messina, Messina, Italy\\
$^{31}$ Dipartimento Interateneo di Fisica `M.~Merlin' and Sezione INFN, Bari, Italy\\
$^{32}$ European Organization for Nuclear Research (CERN), Geneva, Switzerland\\
$^{33}$ Faculty of Electrical Engineering, Mechanical Engineering and Naval Architecture, University of Split, Split, Croatia\\
$^{34}$ Faculty of Engineering and Science, Western Norway University of Applied Sciences, Bergen, Norway\\
$^{35}$ Faculty of Nuclear Sciences and Physical Engineering, Czech Technical University in Prague, Prague, Czech Republic\\
$^{36}$ Faculty of Physics, Sofia University, Sofia, Bulgaria\\
$^{37}$ Faculty of Science, P.J.~\v{S}af\'{a}rik University, Ko\v{s}ice, Slovak Republic\\
$^{38}$ Frankfurt Institute for Advanced Studies, Johann Wolfgang Goethe-Universit\"{a}t Frankfurt, Frankfurt, Germany\\
$^{39}$ Fudan University, Shanghai, China\\
$^{40}$ Gangneung-Wonju National University, Gangneung, Republic of Korea\\
$^{41}$ Gauhati University, Department of Physics, Guwahati, India\\
$^{42}$ Helmholtz-Institut f\"{u}r Strahlen- und Kernphysik, Rheinische Friedrich-Wilhelms-Universit\"{a}t Bonn, Bonn, Germany\\
$^{43}$ Helsinki Institute of Physics (HIP), Helsinki, Finland\\
$^{44}$ High Energy Physics Group,  Universidad Aut\'{o}noma de Puebla, Puebla, Mexico\\
$^{45}$ Horia Hulubei National Institute of Physics and Nuclear Engineering, Bucharest, Romania\\
$^{46}$ Indian Institute of Technology Bombay (IIT), Mumbai, India\\
$^{47}$ Indian Institute of Technology Indore, Indore, India\\
$^{48}$ INFN, Laboratori Nazionali di Frascati, Frascati, Italy\\
$^{49}$ INFN, Sezione di Bari, Bari, Italy\\
$^{50}$ INFN, Sezione di Bologna, Bologna, Italy\\
$^{51}$ INFN, Sezione di Cagliari, Cagliari, Italy\\
$^{52}$ INFN, Sezione di Catania, Catania, Italy\\
$^{53}$ INFN, Sezione di Padova, Padova, Italy\\
$^{54}$ INFN, Sezione di Pavia, Pavia, Italy\\
$^{55}$ INFN, Sezione di Torino, Turin, Italy\\
$^{56}$ INFN, Sezione di Trieste, Trieste, Italy\\
$^{57}$ Inha University, Incheon, Republic of Korea\\
$^{58}$ Institute for Gravitational and Subatomic Physics (GRASP), Utrecht University/Nikhef, Utrecht, Netherlands\\
$^{59}$ Institute of Experimental Physics, Slovak Academy of Sciences, Ko\v{s}ice, Slovak Republic\\
$^{60}$ Institute of Physics, Homi Bhabha National Institute, Bhubaneswar, India\\
$^{61}$ Institute of Physics of the Czech Academy of Sciences, Prague, Czech Republic\\
$^{62}$ Institute of Space Science (ISS), Bucharest, Romania\\
$^{63}$ Institut f\"{u}r Kernphysik, Johann Wolfgang Goethe-Universit\"{a}t Frankfurt, Frankfurt, Germany\\
$^{64}$ Instituto de Ciencias Nucleares, Universidad Nacional Aut\'{o}noma de M\'{e}xico, Mexico City, Mexico\\
$^{65}$ Instituto de F\'{i}sica, Universidade Federal do Rio Grande do Sul (UFRGS), Porto Alegre, Brazil\\
$^{66}$ Instituto de F\'{\i}sica, Universidad Nacional Aut\'{o}noma de M\'{e}xico, Mexico City, Mexico\\
$^{67}$ iThemba LABS, National Research Foundation, Somerset West, South Africa\\
$^{68}$ Jeonbuk National University, Jeonju, Republic of Korea\\
$^{69}$ Johann-Wolfgang-Goethe Universit\"{a}t Frankfurt Institut f\"{u}r Informatik, Fachbereich Informatik und Mathematik, Frankfurt, Germany\\
$^{70}$ Korea Institute of Science and Technology Information, Daejeon, Republic of Korea\\
$^{71}$ KTO Karatay University, Konya, Turkey\\
$^{72}$ Laboratoire de Physique des 2 Infinis, Ir\`{e}ne Joliot-Curie, Orsay, France\\
$^{73}$ Laboratoire de Physique Subatomique et de Cosmologie, Universit\'{e} Grenoble-Alpes, CNRS-IN2P3, Grenoble, France\\
$^{74}$ Lawrence Berkeley National Laboratory, Berkeley, California, United States\\
$^{75}$ Lund University Department of Physics, Division of Particle Physics, Lund, Sweden\\
$^{76}$ Nagasaki Institute of Applied Science, Nagasaki, Japan\\
$^{77}$ Nara Women{'}s University (NWU), Nara, Japan\\
$^{78}$ National and Kapodistrian University of Athens, School of Science, Department of Physics , Athens, Greece\\
$^{79}$ National Centre for Nuclear Research, Warsaw, Poland\\
$^{80}$ National Institute of Science Education and Research, Homi Bhabha National Institute, Jatni, India\\
$^{81}$ National Nuclear Research Center, Baku, Azerbaijan\\
$^{82}$ National Research and Innovation Agency - BRIN, Jakarta, Indonesia\\
$^{83}$ Niels Bohr Institute, University of Copenhagen, Copenhagen, Denmark\\
$^{84}$ Nikhef, National institute for subatomic physics, Amsterdam, Netherlands\\
$^{85}$ Nuclear Physics Group, STFC Daresbury Laboratory, Daresbury, United Kingdom\\
$^{86}$ Nuclear Physics Institute of the Czech Academy of Sciences, Husinec-\v{R}e\v{z}, Czech Republic\\
$^{87}$ Oak Ridge National Laboratory, Oak Ridge, Tennessee, United States\\
$^{88}$ Ohio State University, Columbus, Ohio, United States\\
$^{89}$ Physics department, Faculty of science, University of Zagreb, Zagreb, Croatia\\
$^{90}$ Physics Department, Panjab University, Chandigarh, India\\
$^{91}$ Physics Department, University of Jammu, Jammu, India\\
$^{92}$ Physics Department, University of Rajasthan, Jaipur, India\\
$^{93}$ Physics Program and International Institute for Sustainability with Knotted Chiral Meta Matter (SKCM2), Hiroshima University, Hiroshima, Japan\\
$^{94}$ Physikalisches Institut, Eberhard-Karls-Universit\"{a}t T\"{u}bingen, T\"{u}bingen, Germany\\
$^{95}$ Physikalisches Institut, Ruprecht-Karls-Universit\"{a}t Heidelberg, Heidelberg, Germany\\
$^{96}$ Physik Department, Technische Universit\"{a}t M\"{u}nchen, Munich, Germany\\
$^{97}$ Politecnico di Bari and Sezione INFN, Bari, Italy\\
$^{98}$ Research Division and ExtreMe Matter Institute EMMI, GSI Helmholtzzentrum f\"ur Schwerionenforschung GmbH, Darmstadt, Germany\\
$^{99}$ Saha Institute of Nuclear Physics, Homi Bhabha National Institute, Kolkata, India\\
$^{100}$ School of Physics and Astronomy, University of Birmingham, Birmingham, United Kingdom\\
$^{101}$ Secci\'{o}n F\'{\i}sica, Departamento de Ciencias, Pontificia Universidad Cat\'{o}lica del Per\'{u}, Lima, Peru\\
$^{102}$ Stefan Meyer Institut f\"{u}r Subatomare Physik (SMI), Vienna, Austria\\
$^{103}$ SUBATECH, IMT Atlantique, Nantes Universit\'{e}, CNRS-IN2P3, Nantes, France\\
$^{104}$ Suranaree University of Technology, Nakhon Ratchasima, Thailand\\
$^{105}$ Technical University of Ko\v{s}ice, Ko\v{s}ice, Slovak Republic\\
$^{106}$ The Henryk Niewodniczanski Institute of Nuclear Physics, Polish Academy of Sciences, Cracow, Poland\\
$^{107}$ The University of Texas at Austin, Austin, Texas, United States\\
$^{108}$ Universidad Aut\'{o}noma de Sinaloa, Culiac\'{a}n, Mexico\\
$^{109}$ Universidade de S\~{a}o Paulo (USP), S\~{a}o Paulo, Brazil\\
$^{110}$ Universidade Estadual de Campinas (UNICAMP), Campinas, Brazil\\
$^{111}$ Universidade Federal do ABC, Santo Andre, Brazil\\
$^{112}$ University of Cape Town, Cape Town, South Africa\\
$^{113}$ University of Houston, Houston, Texas, United States\\
$^{114}$ University of Jyv\"{a}skyl\"{a}, Jyv\"{a}skyl\"{a}, Finland\\
$^{115}$ University of Kansas, Lawrence, Kansas, United States\\
$^{116}$ University of Liverpool, Liverpool, United Kingdom\\
$^{117}$ University of Science and Technology of China, Hefei, China\\
$^{118}$ University of South-Eastern Norway, Kongsberg, Norway\\
$^{119}$ University of Tennessee, Knoxville, Tennessee, United States\\
$^{120}$ University of the Witwatersrand, Johannesburg, South Africa\\
$^{121}$ University of Tokyo, Tokyo, Japan\\
$^{122}$ University of Tsukuba, Tsukuba, Japan\\
$^{123}$ University Politehnica of Bucharest, Bucharest, Romania\\
$^{124}$ Universit\'{e} Clermont Auvergne, CNRS/IN2P3, LPC, Clermont-Ferrand, France\\
$^{125}$ Universit\'{e} de Lyon, CNRS/IN2P3, Institut de Physique des 2 Infinis de Lyon, Lyon, France\\
$^{126}$ Universit\'{e} de Strasbourg, CNRS, IPHC UMR 7178, F-67000 Strasbourg, France, Strasbourg, France\\
$^{127}$ Universit\'{e} Paris-Saclay Centre d'Etudes de Saclay (CEA), IRFU, D\'{e}partment de Physique Nucl\'{e}aire (DPhN), Saclay, France\\
$^{128}$ Universit\`{a} degli Studi di Foggia, Foggia, Italy\\
$^{129}$ Universit\`{a} del Piemonte Orientale, Vercelli, Italy\\
$^{130}$ Universit\`{a} di Brescia, Brescia, Italy\\
$^{131}$ Variable Energy Cyclotron Centre, Homi Bhabha National Institute, Kolkata, India\\
$^{132}$ Warsaw University of Technology, Warsaw, Poland\\
$^{133}$ Wayne State University, Detroit, Michigan, United States\\
$^{134}$ Westf\"{a}lische Wilhelms-Universit\"{a}t M\"{u}nster, Institut f\"{u}r Kernphysik, M\"{u}nster, Germany\\
$^{135}$ Wigner Research Centre for Physics, Budapest, Hungary\\
$^{136}$ Yale University, New Haven, Connecticut, United States\\
$^{137}$ Yonsei University, Seoul, Republic of Korea\\
$^{138}$  Zentrum  f\"{u}r Technologie und Transfer (ZTT), Worms, Germany\\
$^{139}$ Affiliated with an institute covered by a cooperation agreement with CERN\\
$^{140}$ Affiliated with an international laboratory covered by a cooperation agreement with CERN.\\

\end{flushleft}

\end{document}